\documentclass[twocolumn, amsmath,amssymb, superscriptaddress]{revtex4}

\usepackage{dcolumn}% Align table columns on decimal point
\usepackage{bm}% bold math
\usepackage{amsmath}
\usepackage{amsfonts}
\usepackage{graphics}
\usepackage[pdftex]{graphicx}
\usepackage{times}
\usepackage{setspace}
\usepackage{color}   % {\color{red} TEXT }
\usepackage{verbatim}
\usepackage[raggedright]{titlesec}

\begin{document}

\title{Statistical Laws Governing Fluctuations in Word Use \\ from Word Birth to Word Death}

\author{Alexander M. Petersen}
\affiliation{Laboratory for the Analysis of Complex Economic Systems, IMT Lucca Institute for Advanced Studies, Lucca 55100, Italy}
%\affiliation{Center for Polymer Studies and Department of Physics, Boston University, Boston, Massachusetts 02215, USA}\
\author{Joel Tenenbaum}
\affiliation{Center for Polymer Studies and Department of Physics, Boston University, Boston, Massachusetts 02215, USA}
\author{Shlomo Havlin}
\affiliation{Minerva Center and Department of Physics, Bar-Ilan University, Ramat-Gan 52900, Israel}
\author{H. Eugene Stanley}
\affiliation{Center for Polymer Studies and Department of Physics, Boston University, Boston, Massachusetts 02215, USA}
\begin{abstract} 
%The growth in the use of a given word provides  insight into the coevolution of language and culture. 
%Since little is known about the aggregate dynamics of language across time or across language, w
We analyze the dynamic properties of $10^{7}$ words  recorded in English, Spanish and Hebrew over the period 1800--2008 in order  to gain insight into the coevolution of language and culture. We report language independent  patterns  useful as benchmarks for theoretical models of language evolution. A significantly decreasing (increasing) trend in the  birth (death) rate of words indicates a recent shift in the selection laws governing word use. For new words, we
observe a peak in the growth-rate fluctuations around 40 years after introduction, consistent with the typical entry time into standard dictionaries and  the human generational timescale.
Pronounced changes in the dynamics of language during periods of war shows that word correlations, occurring across time and between words, are largely influenced by coevolutionary social, technological, and political factors. We quantify cultural memory by analyzing the long-term correlations in the use of individual words using detrended fluctuation analysis.
%Similarities between the growth dynamics of individual words and individual economic entities suggests that both cooperation and competition are governed by a common evolutionary mechanism. 
% currently 150 words
\end{abstract}
\date{\today}
%150 word abstract
%The Methods section is limited to 1500 words. Figure legends are limited to 350 words. References are limited to 60. Footnotes are not used.
%Depending on the word count, Articles may have up to 8 display items (figures and/or tables).

\maketitle

\footnotetext[1]{ Corresponding author: Alexander M. Petersen \\
{\it E-mail}: \text{petersen.xander@gmail.com}
}
%\section{Introduction} 

Statistical laws describing the  properties of word use, such as Zipf's law \cite{Zipf, Zipfrecent, PlosOneZipf, 2regimeZipf,variationZipf, ScalingZipf} and Heaps' law \cite{HeapsOrig, MetaHeaps}, have been thoroughly tested and
modeled. These statistical laws are based on static snapshots of written language using empirical data aggregated over relatively 
small time periods and comprised of relatively small corpora ranging in size  from 
individual texts \cite{Zipf, Zipfrecent} to relatively small collections of topical texts \cite{PlosOneZipf,2regimeZipf}. 
However, language is a fundamentally dynamic complex system, consisting of heterogenous entities at the level of the units (words) and the interacting users (us).
Hence, we begin this paper with two questions: (i) Do languages exhibit dynamical patterns? (ii) Do individual words exhibit dynamical patterns? 

The coevolutionary nature of language requires analysis both at the macro and micro scale. Here we apply interdisciplinary concepts to
empirical language data collected in a massive book digitization effort by {\it Google Inc.}, which  recently unveiled a database of words in seven languages, after
having  scanned approximately 4\% of the world's books. 
The massive ``n-gram''
project \cite{googledata} allows for a  novel view into the growth dynamics of word use and the
birth and death processes of words in accordance with evolutionary selection laws \cite{EvolLang}.

A recent analysis  of this database by Michel et al. \cite{googlepaper} addresses numerous well-posed
questions rooted in cultural anthropology using case studies of individual words.  Here we take an alternative approach by analyzing  the {\it aggregate} properties of the language dynamics recorded in  the {\it Google Inc.}  data in a systematic way, using the word counts of  every word recorded over the 209-year time period  1800 -- 2008 in the English, Spanish, and Hebrew text corpora.  This period spans the incredibly rich cultural history that includes several international wars, revolutions, and numerous technological paradigm shifts. Together, the data comprise over $1 \times 10^{7}$ distinct words. We use 
 concepts from economics to gain quantitative insights into the role of exogenous 
factors on the evolution of language, combined with methods from statistical physics to quantify 
the competition arising from correlations between words \cite{WordnetlexiconHierarchy,SemanticNetwork,textHeirarchyCorr} and the memory-driven autocorrelations in $u_{i}(t)$ across time
\cite{FractalCorrCorpora,ReturnIntervalsLanguage,WordBursts}.

For each corpora comprising millions of distinct words, we use a general word-count framework which accounts for the underlying growth of language over time.
We first define the quantity $u_{i}(t)$ as the number  of uses of word $i$ in year $t$.  
Since the number of books and the number of distinct words have grown dramatically over time, we define the {\it relative} word use, $f_{i}(t)$,  as the fraction  of uses of word $i$ out of all  word  uses in the same year, 
\begin{equation}
f_{i}(t) \equiv u_{i}(t)/N_{u}(t) \ ,
\end{equation}
where the quantity $N_{u}(t) 
\equiv \sum_{i=1}^{N_{w}(t)} u_{i}(t)$ is the total number of indistinct word  uses digitized from books printed in year $t$
and $N_{w}(t)$ is the total number of distinct words digitized from books printed in year $t$.
To quantify the dynamic properties of word prevalence at the micro scale and their relation to socio-political factors at the macro scale, we analyze the logarithmic growth rate commonly used in  finance and  economics,
\begin{eqnarray}
r_{i}(t) &\equiv& \ln f_{i}(t+\Delta t)-\ln f_{i}(t) = \ln \Big( \frac{f_{i}(t+\Delta t)}{f_{i}(t)}\Big) \ .
%&=& g_{i}(t)  - \ln [N_{u}(t)/N_{u}(t-\Delta t)] \ .
\label{r2}
\end{eqnarray}

The relative use $f_{i}(t)$  depends on the intrinsic grammatical utility of the word (related to the number of ``proper''
sentences that can be constructed using the word), the semantic utility of the word (related to the number of meanings a given
word can convey), and other idiosyncratic details related to topical context. 
Neutral null models for the evolution of language  define the relative use of a word as its {\it ``fitness''} \cite{LanguageNeutralEvol}. In such models, the word frequency is the only factor determining the survival capacity of a word. 
In reality, word competition depends on more subtle features of language, such as  the cognitive aspects of efficient  communication. 
For example, the emergence of robust categorical naming patterns observed across many cultures is  regarded to be the result of complex discrimination tactics 
shared by intelligent communicators.
This is evident in the finite set of words describing the continuous spectrum of
  color names, emotional states, and other categorical  sets  \cite{StatPhysLangDyn,ColorNaming,EmergenceTopicality}. 

In our analysis we treat words with equivalent meanings but with different spellings (e.g. color versus colour) as distinct words, since  we view the  competition among synonyms and alternative spellings in the linguistic arena as a key ingredient in
complex evolutionary dynamics \cite{NowakLanguage,EvolLang}. 
For instance, with the advent of automatic spell-checkers 
in the digital era, words recognized by  spell-checkers  receive a significant boost in their ``reproductive fitness'' at the
 expense of their misspelled or unstandardized counterparts. 

  \begin{figure}
\centering{\includegraphics[width=0.48\textwidth]{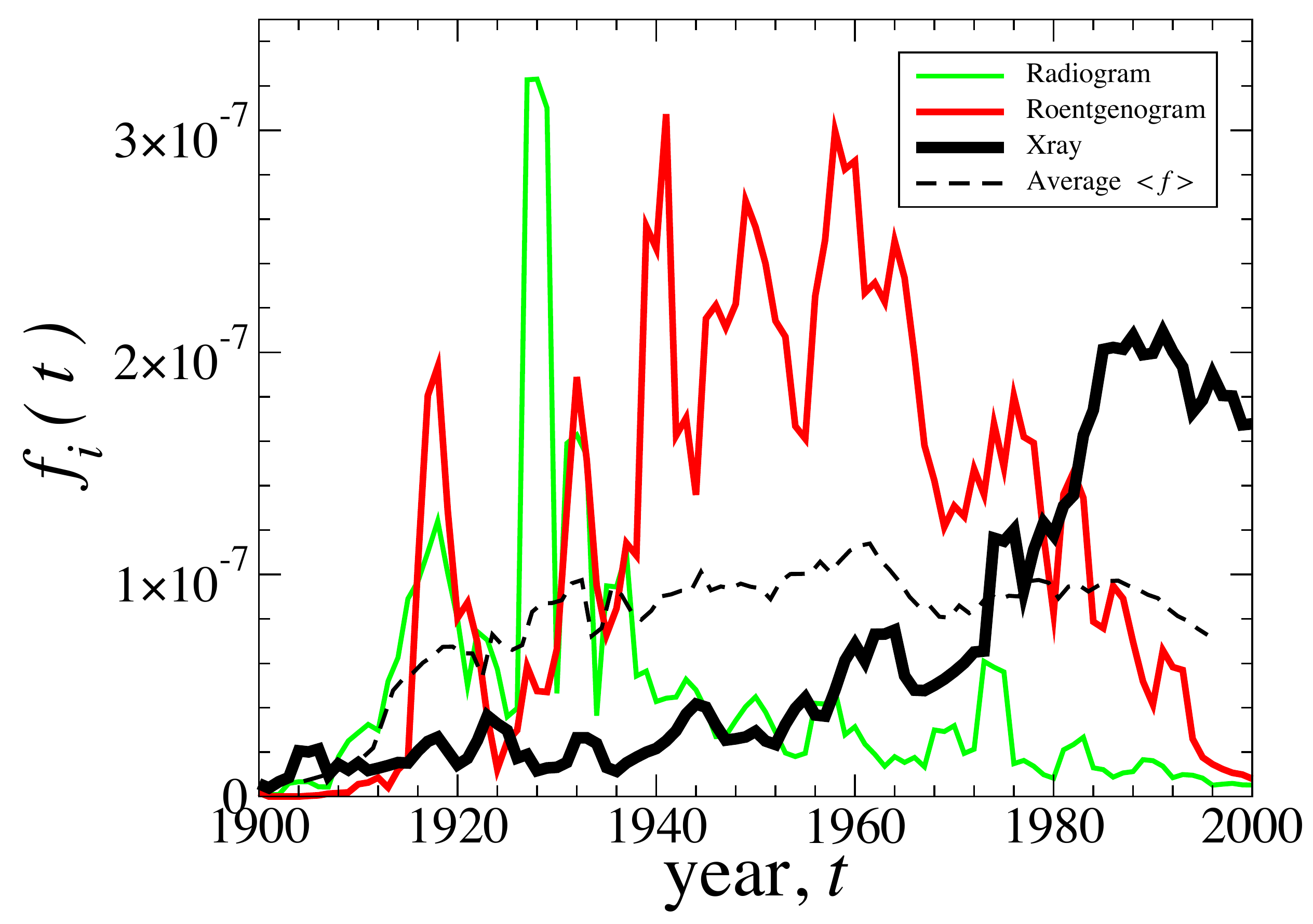}}
  \caption{  {\bf Word extinction.} The  English word ``Roentgenogram'' derives from the 
 Nobel prize winning scientist and discoverer of the x-ray, Wilhelm R\"{o}ntgen (1845--1923). The prevalence of this word was quickly  challenged by two main competitors, ``X-ray'' (recorded as ``Xray'' in the database) and ``Radiogram.''  The arithmetic mean frequency  of these three time series is relatively constant over the 80-year period 1920--2000, $\langle f \rangle \approx 10^{-7}$,  illustrating the  limited linguistic ``market share'' that can be achieved by any competitor. We conjecture that the main reason ``Xray'' has a higher frequency  is due to the ``fitness gain'' from its efficient short word length and also due to the fact that English has become the base language for scientific publication.} 
\label{radiology}
\end{figure}

In the linguistic arena, not just ``defective'' words die, even significantly used words can become extinct.  Fig. \ref{radiology} shows  three once-significant  words: ``Radiogram,'' ``Roentgenogram,'' and ``Xray''. These words compete  for the majority share of nouns referring to what is now commonly known as an ``X-ray'' (note that such dashes are discarded in Google's digitization  process). 
The  word ``Roentgenogram'' has since become extinct, even though 
it was the most common term for several decades in the 20th century.  It is likely that two main factors -- (i) communication and information efficiency bias toward the use of shorter words \cite{efficientcommunication} and (ii) the adoption of English as the leading global language for science -- secured the eventual success of the word ``Xray'' by the year 1980. It goes without saying that there are many social and technological factors driving language change.

We begin this paper by analyzing the vocabulary growth of each language over time. We then analyze the lifetime growth trajectories of the set of words that are new
to each language  to gain quantitative insight into ``infant'' and ``adult''  stages of  individual words. Using two sets of words, (i) the relatively new words, and (ii) the most common words, we analyze  the statistical properties of word growth. Specifically, we calculate  the probability density function $P(r)$  of growth rate $r$ and calculate the size-dependence  of the standard deviation $\sigma(r)$ of growth rates. In order to gain insight into the long-term  cultural memory, we conclude the analysis by measuring the autocorrelations in word use by applying detrended fluctuation analysis (DFA) to individual time series.  

\section*{Results}
\noindent{\bf Quantifying the birth rate and the death rate of words.} Just as a new species can be born into an environment, a word can emerge in a language. Evolutionary selection
laws can apply pressure on the sustainability of new words since there are  limited resources (topics, books, etc.) for the use of
words. Along the same lines, old words can be driven to extinction when cultural and technological factors limit the use of a word, 
in analogy to the environmental factors that can 
change the survival capacity of a living species by altering its ability to survive and reproduce.

 \begin{figure}
\centering{\includegraphics[width=0.48\textwidth]{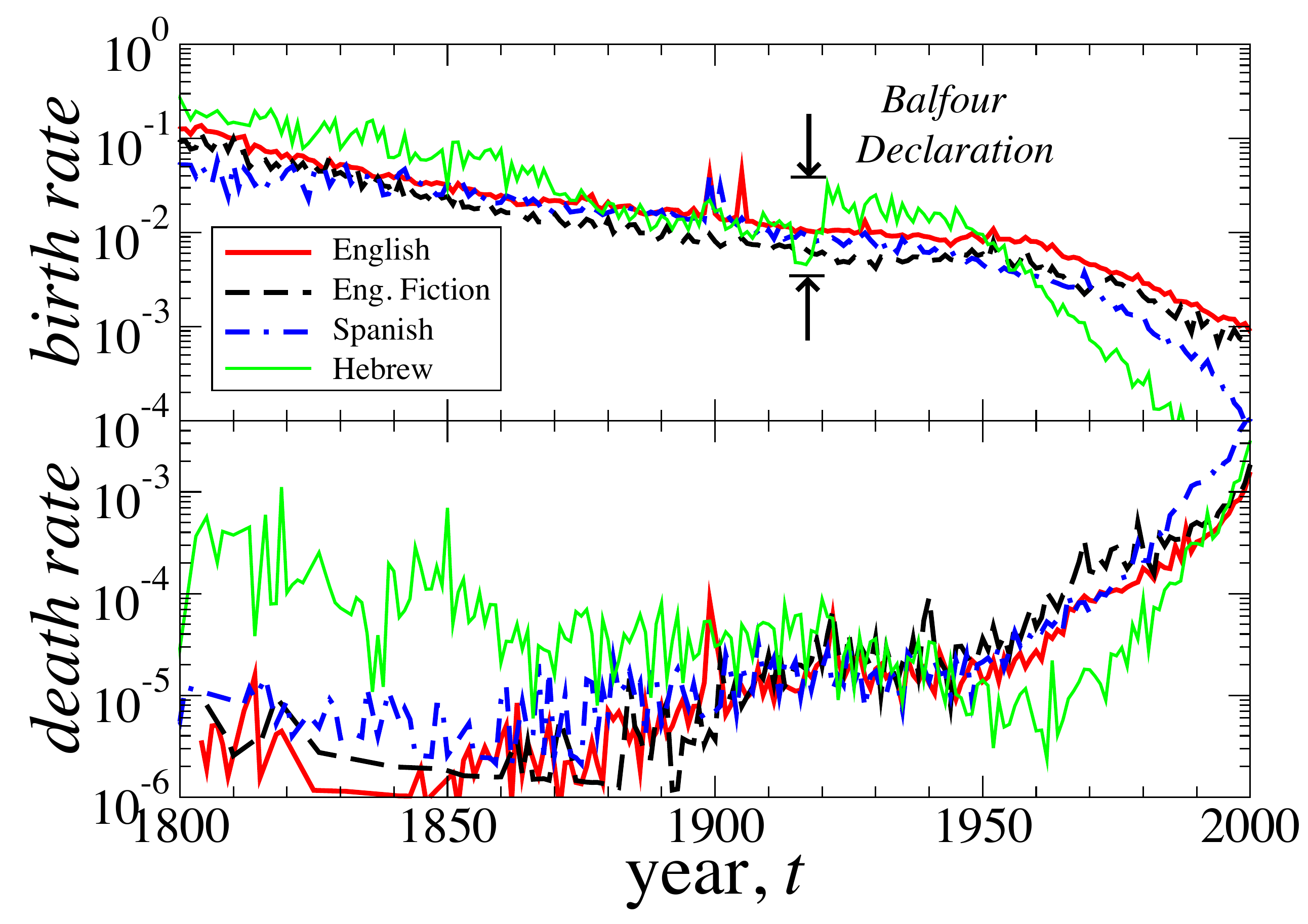}}
  \caption{ {\bf Dramatic shift in the birth rate and death rate of words. } The word birth rate $\gamma_{b}(t)$ and the word death rate $\gamma_{d}(t)$  show marked underlying changes in word use competition which affects the entry rate and the sustainability of existing words.
 The modern print era shows a marked increase in the death rate of words which likely correspond to  low fitness, misspelled and (technologically) outdated words. A simultaneous decrease in the birth rate of new words is consistent with the decreasing marginal need for new words indicated by the sub-linear  allometric scaling between vocabulary size and total corpus size (Heaps' law) \cite{LanguageAll}.  Interestingly, we  quantitatively observe the impact of the Balfour Declaration in 1917, the circumstances surrounding which effectively rejuvenated  Hebrew as a national language, resulting in a 5-fold increase in the birth rate of  words in the Hebrew corpus.} 
\label{bdrate}
\end{figure}

\begin{figure}
\centering{\includegraphics[width=0.48\textwidth]{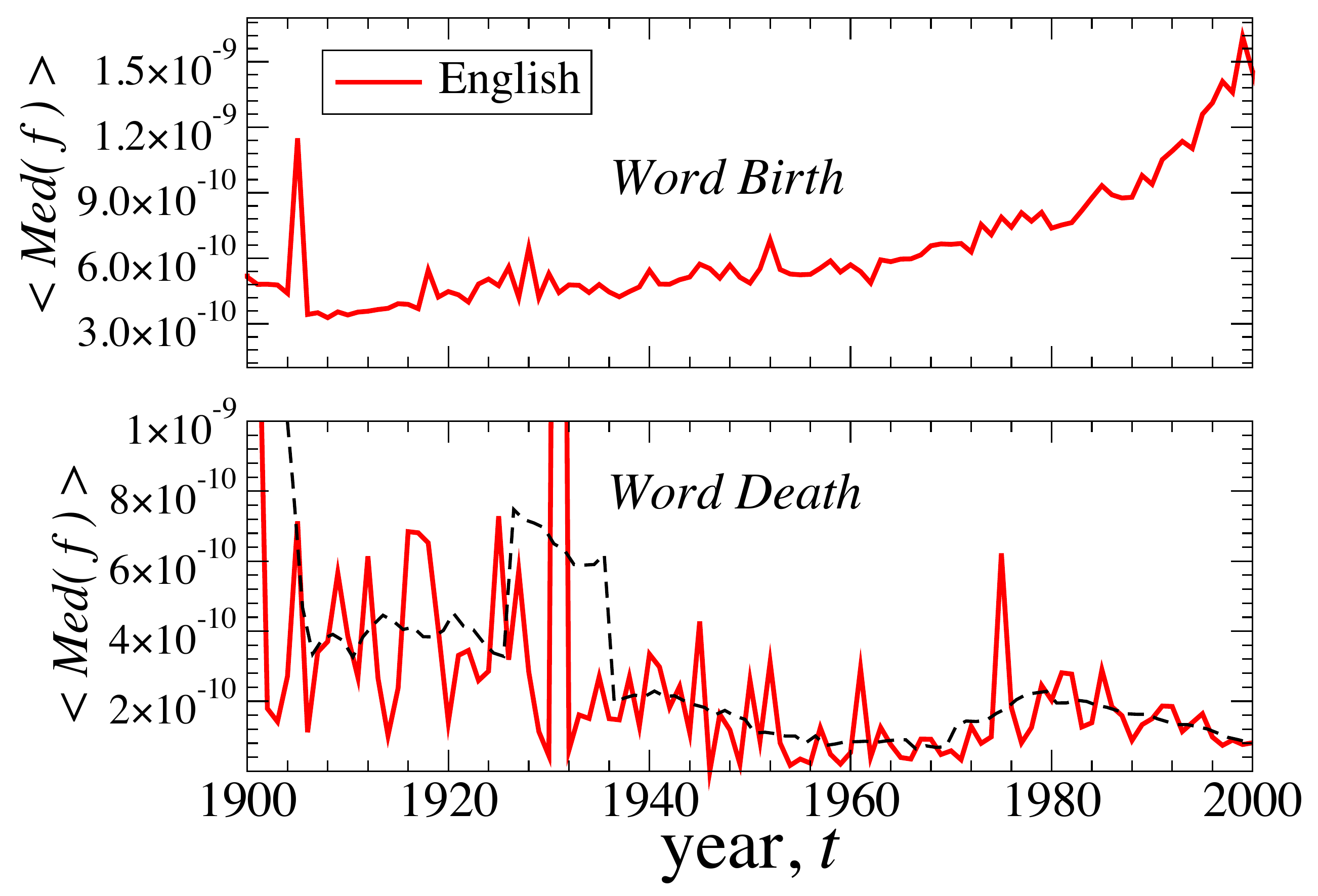}}
  \caption{ {\bf Survival of the fittest in the entry process of words.} Trends in the relative uses of words that either were born or died in a given year  show that the entry-exit forces  largely depend on the relative use of the word. 
  For the English corpus, we calculate the average of the median lifetime relative use, $\langle \text{Med}(f_{i}) \rangle$, for all words born in year $t$ (top panel) and 
  for all words that died in year $t$ (bottom panel), which shows a 5-year moving average (dashed black line).
  There is a dramatic increase in the  relative use (``utility'') of  newborn words over the last 20--30 years, 
  likely corresponding to new technical terms, which are necessary for the communication of core modern technology and ideas.
  Conversely, with higher editorial standards and the recent use of word processors which include spelling standardization technology, the words that are dying are those words with low relative use.  We confirm by visual inspection that the lists of dying words contain mostly misspelled and nonsensical words. } 
\label{BDAveMedUse}
\end{figure}

We define the birth year $y_{0,i}$ as 
the year $t$ corresponding to the first instance of $f_{i}(t) \geq 0.05 f^{m}_{i}$, where $f^{m}_{i}$  is median word use $f^{m}_{i} = Median\{u_{i}(t)\}$ of a given word over its recorded lifetime in the {\it Google} database.
  Similarly, we define the death year $y_{f,i}$ as the last year $t$
during which the word use satisfies $f_{i}(t) \geq 0.05 f^{m}_{i}$. 
We use the relative word use threshold $0.05 f^{m}_{i}$ in order
to avoid anomalies arising from extreme fluctuations in $f_{i}(t)$ over the lifetime of the word. 
The results obtained using  threshold $0.10 f^{m}_{i}$ did not show a significant qualitative difference.

The significance of word births $\Delta_{b}(t)$ and word deaths $\Delta_{d}(t)$ for each year $t$ is related to the vocabulary size $N_{w}(t)$ of a given language.  
We define the birth rate $\gamma_{b}$ and death rate $\gamma_{d}$ by normalizing the number of births and deaths in a given year $t$  to the total number of distinct words $N_{w}(t)$ recorded in the same year $t$, so that
\begin{eqnarray}
\gamma_{b}(t) \equiv \Delta_{b}(t) / N_{w}(t)  \ , \\ \nonumber
\gamma_{d}(t) \equiv \Delta_{d}(t) / N_{w}(t) \ .
\end{eqnarray}
This definition yields a proxy for the rate of emergence and disappearance of words. We restrict our analysis to words with birth-death duration $y_{f,i}-y_{0,i}+1 \geq 2$ years and  to words with first recorded use
 $t_{0,i} \geq 1700$, which selects for relatively new words in the history of a language.

 The $\gamma_{b}(t)$ and $\gamma_{d}(t)$ time series plotted in Fig. \ref{bdrate}  for the 200-year period 1800--2000 show  trends that intensifies after the 1950s. 
The modern era of publishing, which is characterized by more strict editing procedures at publishing houses,
computerized word editing and automatic spell-checking technology,  shows a drastic increase in the death rate of words.
%We do not show the calculations for most recent years since 2000, which show continued trends, since the numbers  may be affected by insignificant short-lived words.
 Using visual inspection we verify  most changes to the vocabulary  in the last 10--20 years are  due to the extinction of misspelled words and  nonsensical print errors, and to the decreased 
 birth rate of new misspelled variations and genuinely new words. This phenomenon reflects  the decreasing marginal need for new words, consistent with the sub-linear Heaps' law observed for all Google 1-gram corpora  in \cite{LanguageAll}. 
Moreover, Fig. \ref{BDAveMedUse} shows that $\gamma_{b}(t)$ is largely comprised of  words with relatively large median $f_{c}$ while $\gamma_{d}(t)$  
is almost entirely comprised of words with relatively small median $f_{c}$ (see also Fig.  \ref{BDRateFc} in the Supplementary Information (SI) text). 
Thus, the new words of tomorrow are likely be core words that are widely used.

We note that the main source of error in the calculation of birth and death rates are  OCR (optical character recognition) errors in the digitization process, which could be responsible for a significant fraction of misspelled and nonsensical words existing in the data. An additional source of error is the variety of orthographic properties of language that can make very subtle variations of words, for example through the use of hyphens and capitalization,  appear as distinct words when applying OCR.
The digitization of many books in the computer era does not require OCR transfer, since the manuscripts are themselves digital,  and so there may be a bias resulting from this recent paradigm shift. We confirm that the statistical patterns found using  post 2000- data  are consistent with the patterns that extend back several hundred years \cite{LanguageAll}.
 
 \begin{figure*}
\centering{\includegraphics[width=0.70\textwidth]{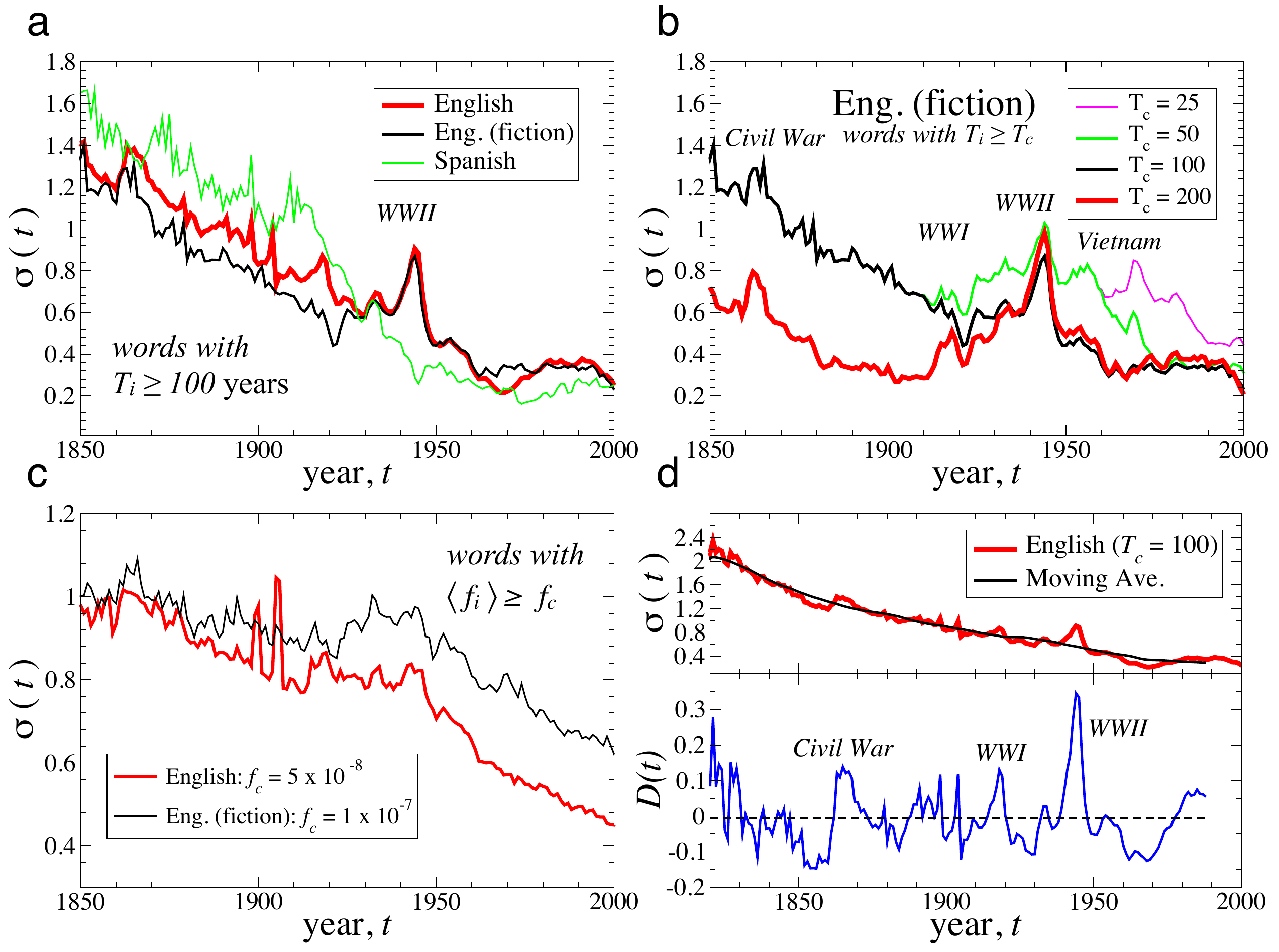}}
  \caption{ {\bf The significance of historical events on the evolution of language.}
  The standard deviation  $\sigma(t)$ of growth rates 
demonstrates the sensitivity of language to international events (e.g. World War II). 
For all languages there is an overall 
decreasing trend in $\sigma(t)$ over the  period 1850--2000.
However, the increase in $\sigma(t)$ during WWII
 represents a``globalization'' effect, whereby
societies are brought 
together by a common event and a unified media. 
Such contact between relatively isolated systems necessarily leads to information flow, much as in the case of thermodynamic
heat flow between two systems, initially at different temperatures, which are then  brought into contact.
{\bf (a)}  $\sigma(t)$ calculated for the relatively new words
with $T_{i} \geq 100$ years. The Spanish
corpus does not 
show an increase in $\sigma(t)$ during  World War II, indicative of the 
  relative isolation of South America and Spain from the European conflict. {\bf (b)}   $\sigma(t)$ for 4 sets of
relatively new words that
meet the criteria $T_{i} \geq T_{c}$  and $t_{i,0} \geq 1800$. The oldest ``new'' words ($T_{c}=200$) demonstrate the 
most significant increase in $\sigma(t)$ during  World War II, with a peak around 1945. {\bf (c)} The  standard deviation $\sigma(t)$ for the most common words is decreasing with time, suggesting that they have saturated and are being ``crowding out'' by new competitors. This set of words meets the criterion that the average relative use exceeds a threshold, $\langle f_{i} \rangle \geq
f_{c}$, which we define for each corpus. 
{\bf (d)} We compare the variation  $\sigma(t)$ for relatively new English words, using $T_{i} \geq 100$, with the 20-year
moving average over the time period 1820--1988. The deviations show that  $\sigma(t)$ increases abruptly during times of conflict, such as the American Civil War
(1861--1865), World War I 
(1914--1918) and World War II (1939--1945), and also during the 1980s and 1990s, possibly as a result of new
digital media (e.g. the 
internet) which offer new environments for the evolutionary dynamics of word use. $D(t)$ is the difference between the moving average and $\sigma(t)$.
  }
\label{Ysigmar}
\end{figure*}

Complementary to the death of old words is the birth of new words, which are commonly associated with new social and technological trends. Topical words in  media  can display long-term persistence patterns analogous to earthquake shocks  \cite{blogOmori, BlogWordDynamics}, and can result in a 
new word having larger fitness than
related ``out-of-date'' words (e.g. blog vs. log, email vs. memo).
Here we show that a comparison of the growth dynamics between different languages can also
illustrate the local cultural factors  that influence different regions of the world. 
Fig. \ref{Ysigmar} shows how international crisis can lead to globalization of language
 through common  media attention and increased lexical diffusion.  Notably, as illustrated in Fig. \ref{Ysigmar}(a), we find that international conflict only
perturbed the participating  languages,  while minimally affecting the languages of the 
nonparticipating regions, e.g. the Spanish speaking countries during WWII. \\
%We note that the English corpus and the Spanish corpus are the collections of literature from
%several nations, whereas the 
%Hebrew corpus is more localized.

\noindent{\bf The lifetime trajectory of words.} Between birth and death, one contends with the interesting question of how the use of words evolve when they are ``alive.''
We focus our efforts toward quantifying the relative change in word use over time, both over the word
lifetime and 
throughout the course of history. In order to analyze separately these two time frames, we select  two sets of
words: (i) relatively new 
words with ``birth year'' $t_{0,i}$ later  than 1800, so that the relative age  $\tau \equiv t -  t_{0,i}$ of word $i$ is the
number of years after the word's first 
occurrence in the database, and (ii) relatively common words, typically with $ t_{0,i} <$ 1800. 

We analyze dataset (i) words (summary statistics in Table \ref{TableSummary2}) so that we can control for properties of the
growth dynamics that
 are related to the various stages of a word's life trajectory (e.g. an ``infant'' phase, an ``adolescent'' phase, and a ``mature''
phase).  
For comparison with the young words, we also analyze the growth rates of dataset (ii) words  in the next section (summary statistics in Table \ref{TableSummary3}). These words   are presumably old enough that they are in a stable mature phase. 
 We select dataset (ii) words using the criterion  $\langle f_{i} \rangle \geq
f_{c}$, where  $\langle f_{i} \rangle = \sum_{\tau =1}^{T_{i}}f_{i}(\tau) / T_{i}$ is 
the average relative use of the word $i$ over the word's lifetime $T_{i}= t_{0,f}-t_{0,i}+1$, and $f_{c}$ is a cutoff threshold derived form the Zipf rank-frequency distribution \cite{Zipf} calculated for each corpus \cite{LanguageAll}. 
In Table
\ref{TableSummary} we summarize    the entire data for the 209-year period 1800--2008 for each of the four {\it Google}  language sets analyzed. 

\begin{figure}
\centering{\includegraphics[width=0.45\textwidth]{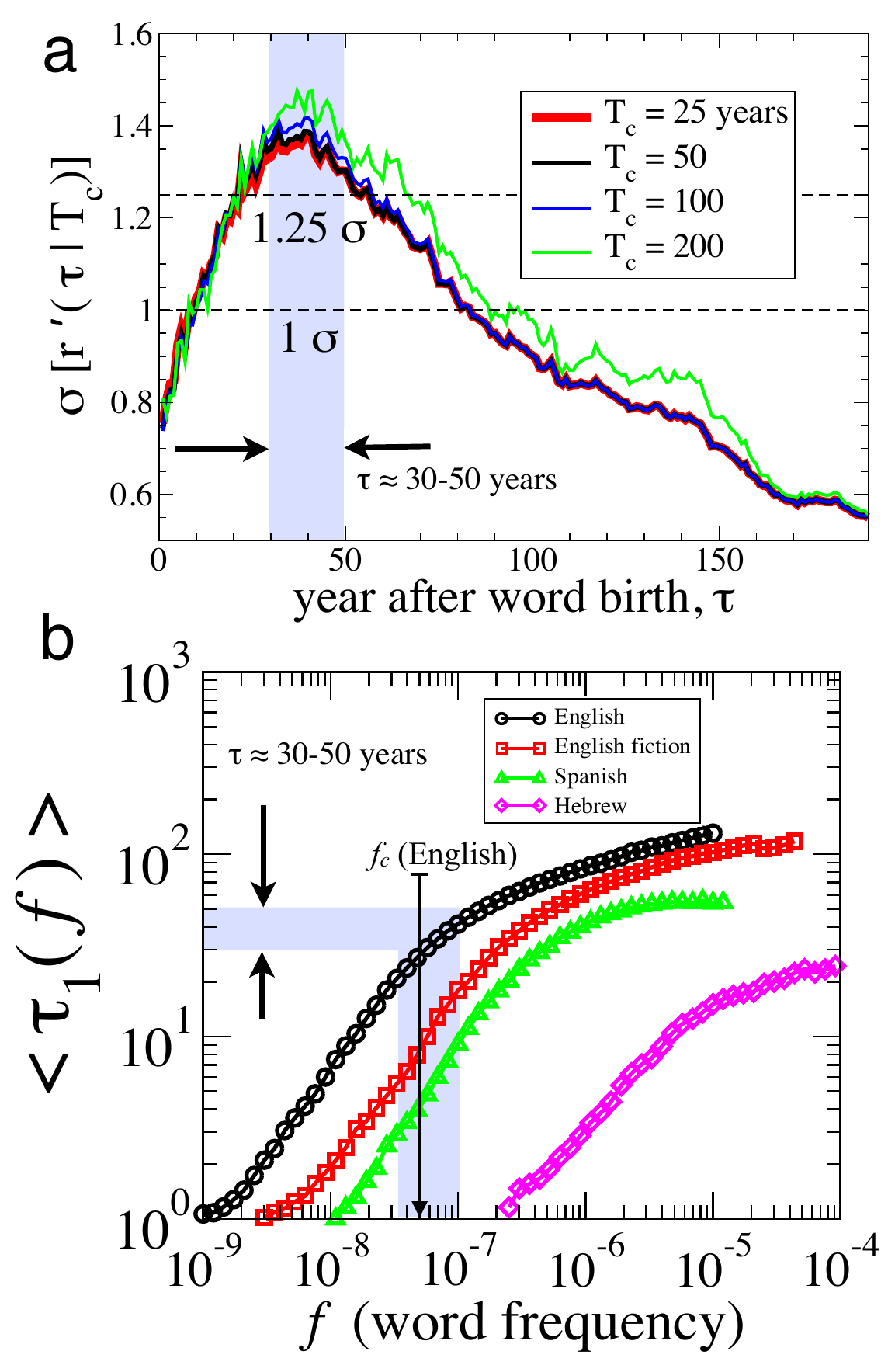}}
  \caption{ {\bf Quantifying the tipping point for word use.} {\bf (a)} The maximum in the standard deviation $\sigma$ of growth rates during the ``adolescent'' period $\tau \approx$ 30--50  indicates 
  the characteristic time scale for words being incorporated into the standard lexicon, i.e. 
  inclusion in popular dictionaries. In Fig. \ref{AveSDTraj} we plot the  average growth rate trajectory $\langle
r'(\tau | T_{c}) \rangle$ which 
  shows relatively large positive growth rates during approximately the same 20-year period. {\bf (b)} 
  The first passage time $\tau_{1}$  \cite{FirstPassage}  is defined as the number years for the relative use of a new word $i$ to exceed a given $f$-value  for the first time,  $f_{i}(\tau_{1}) \geq  f$.   For relatively new words with $T_{i}\geq 100$ years we calculate the average first-passage time $\langle
\tau_{1}(f) \rangle$  for  a large range of $f$.  We estimate for each language the $f_{c}$ representing the threshold for a word belonging to the standard ``kernel'' lexicon \cite{2regimeZipf}. This method demonstrates that the English corpus threshold $f_{c} \equiv 5 \times 10^{-8}$  maps to the first passage time corresponding to the peak period $\tau \approx 30-50$ years in $\sigma(\tau)$ shown in panel (a).  } 
\label{SDtrajectory}
\end{figure}

Modern words typically are  born in relation to technological or cultural events, e.g. ``Antibiotics.'' We ask if there exists a characteristic time  for a word's general acceptance. 
In order to search for patterns in the growth rates as a function of relative word age, for each new  word $i$ at its age $\tau$, we analyze the ``use trajectory'' $f_{i}(\tau)$ and the ``growth rate trajectory'' $r_{i}(\tau)$. 
 So that we may combine the individual trajectories of words of varying prevalence, we normalize  each  
 $f_{i}(\tau)$  by its average $\langle f_{i} \rangle$, obtaining
a normalized use 
trajectory $f'_{i}(\tau) \equiv f_{i}(\tau) / \langle f_{i} \rangle$. We perform an analogous normalization procedure
for each  $r_{i}(\tau)$, normalizing instead by the growth rate 
standard deviation $\sigma [r_{i}]$,  so that $r'_{i}(\tau) \equiv r_{i}(\tau) / \sigma [r_{i}]$ (see the Methods section for further detailed description). 

Since some words will die and other words will increase in use as a result of the standardization of language, we hypothesize that the  average growth rate trajectory will show large fluctuations around the time scale for the
transition  of a
word into regular use. In order to quantify this transition time scale, we create a subset $\{i\ | T_{c}\}$ of word trajectories $i$ by combining 
words that meets an age criteria  $T_{i} \geq T_{c}$. Thus, $T_{c}$ is a threshold to distinguish words that were born in different historical
eras and which have varying longevity. For the values $T_{c} = 25, 50, 100,$ and 200 years, we select all words that
have a lifetime longer than $T_{c}$ and calculate the average and standard 
deviation for each set of  growth rate trajectories as a function of word age $\tau$. 

In Fig. \ref{SDtrajectory} we plot
$\sigma[r'_{i}(\tau | T_{c})]$ for the English corpus, which shows a broad peak  around $\tau_{c} \approx$ 30--50 years for each $T_{c}$ subset
before the fluctuations saturate after the word enters a stable growth phase. 
A similar peak is observed for each corpus analyzed (Figs. \ref{AveSDTraj}--\ref{AveSDTrajHeb}).
This single-peak growth trajectory is consistent with theoretical models for logistic spreading and the fixation of words in a population of learners \cite{LangEcophysics}.
Also, since we weight the average according to $\langle f_{i} \rangle$, the time scale $\tau_{c}$  is likely associated with 
the characteristic time for a new word  to reach sufficiently wide acceptance that the word is included in a typical
 dictionary. 
  We note that this time scale is close to the generational time scale for humans, corroborating evidence that languages require only one generation to 
  drastically evolve  \cite{LangEcophysics}.
  \\
 
\begin{figure}
\centering{\includegraphics[width=0.45\textwidth]{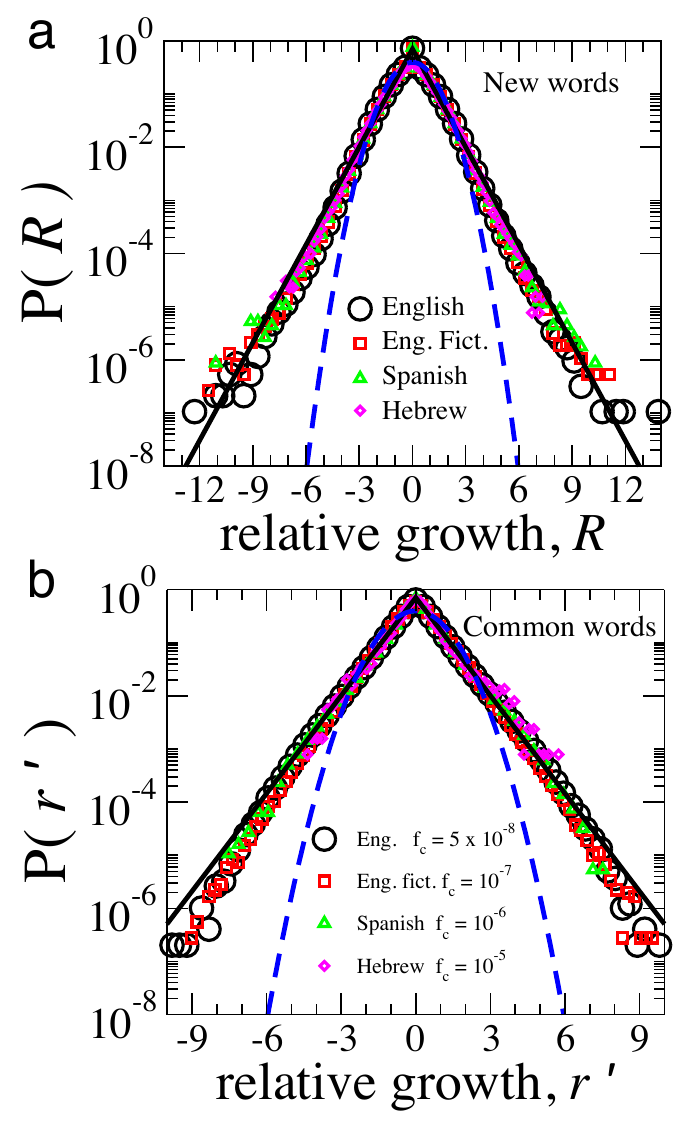}}
  \caption{  {\bf Common leptokurtic growth distribution for new words and common words.}  {\bf (a)} 
 Independent of language, the growth rates of relatively new words are distributed according to the  Laplace distribution centered around $R \approx 0$ defined in Eq. (\ref{LaplaceEqn}). 
 The the growth rate $R$ defined in Eq. (\ref{R}) is measured in units of standard deviation, and accounts for age-dependent and word-dependent factors. Yet, even with these normalizations, we still observe an excess number of  $\vert R \vert \geq 3 \sigma$ events.
This fact is demonstrated by the leptokurtic form of each $P(R)$, which exhibit the excess tail frequencies when compared with a unit-variance Gaussian distribution (dashed blue curve). The Gaussian distribution is the predicted
distribution for the  Gibrat proportional growth model, which is a candidate neutral null-model  for the growth dynamics of word use
\cite{Growth6}. 
  The prevalence of large growth rates illustrate the possibility that words can have large variations in use even over the course of a year.  
The growth variations are intrinsically related to the dynamics of everyday life and reflect the cultural and technological shocks in society. We analyze word use data over the 
  time period 1800-2008 for new words $i$ with  lifetimes  $T_{i} \geq T_{c}$, where we show data calculated for $T_{c}=100$
years. 
 {\bf (b)} PDF $P(r')$ of the annual relative growth rate $r'$ for  all words which satisfy $\langle f_{i} \rangle \geq f_{c}$ (dataset \#ii words which are relatively common words). 
In order to select relatively frequently used words, we use the following criteria:
$T_{i} \geq 10$ years, $1800 \leq t \leq 2008$, and $\langle f_ {i} \rangle \geq f_{c}$. The growth rate $r'$ does not account for age-dependent factors  since the common words are  likely in the mature phase of their lifetime trajectory.
In each panel, we plot a Laplace distribution with unit variance (solid black lines) and the Gaussian distribution with unit variance
(dashed blue curve) for reference. 
} 
\label{dRUPDF}
\end{figure}

\noindent{\bf Empirical laws quantifying the growth rate distribution.} How much do   the growth rates vary from word to word? The answer to  this question can help distinguish between candidate models for the evolution of word utility. Hence, we calculate the probability density
function (pdf) of $R \equiv r'_{i}(\tau) / \sigma[r'(\tau | T_{c})]$. Using this quantity accounts for the fact that we are aggregating  growth rates
of words of varying ages. The empirical pdf $P(R)$ shown in Fig. \ref{dRUPDF} is leptokurtic and remarkably symmetric  around
$R\approx 0$. These empirical facts are also observed in studies of  the growth rates of economic institutions  \cite{Growth1b, Growth6, Growth1,   GrowthEcon}. Since the $R$ values are normalized and
detrended according to  the age-dependent standard deviation $\sigma[r'(\tau | T_{c})]$, the standard deviation   is
 $\sigma(R) =1$ by construction.

A candidate model for the growth rates of word use is the Gibrat proportional growth process \cite{Growth1,Growth6},  which predicts a Gaussian distribution for $P(R)$.  However, we observe the  ``tent-shaped'' pdf $P(R)$ which is well-approximated by a  Laplace  (double-exponential) distribution, defined as
\begin{equation}
P(R) \equiv \frac{1}{\sqrt{2} \sigma(R)} \exp[-\sqrt{2} \vert R - \langle R \rangle \vert /\sigma(R)] \ . 
\label{LaplaceEqn}
\end{equation}
Here the average growth rate $\langle R \rangle$ has two properties: (a) $\langle R \rangle \approx 0$ and (b) $\langle R \rangle \ll
\sigma(R)$. Property (a) arises from the fact that the growth rate of distinct words is quite small  on the annual basis (the growth rate of books in the Google English database is $\gamma_{w} \approx
0.011$  \cite{LanguageAll}) and property (b) arises from the fact that $R$ is defined in units of standard deviation.
 Being leptokurtic, the Laplace distribution  predicts  an excess number of  events $>3\sigma$ as compared to the Gaussian
distribution. For example, comparing the likelihood 
of events above the $3\sigma$ event threshold,  the Laplace distribution displays  a five-fold excess in the probability $P(\vert R- \langle R
\rangle \vert> 3\sigma)$, where  $P(\vert R - \langle R \rangle \vert> 3\sigma) = \exp[-3\sqrt{2}] \approx  0.014$ for
the Laplace distribution, whereas $P(\vert R-
\langle R \rangle \vert> 3\sigma) = \text{Erfc}[3 / \sqrt{2}] \approx  0.0027$ for the Gaussian distribution. The
large $R$ values correspond to periods of rapid growth and decline in the use of words during the crucial ``infant'' 
and ``adolescent'' lifetime phases. In  Fig.
\ref{dRUPDF}(b) we also show that the growth rate distribution $P(r')$ for the relatively common words comprising dataset (ii) is also well-described by the Laplace distribution. 

For hierarchical systems consisting of units each with
complex internal structure \cite{Growth2} (e.g. a given country consists of industries,  each of which consists of companies, each of which consists of
internal subunits), a non-trivial scaling relation  between the standard deviation of growth rates $\sigma(r|S)$ and the system size $S$ has the form 
\begin{equation}
\sigma(r | S_{i}) \sim S_{i}^{-\beta} \ .
\label{sigmaRS}
\end{equation}
 The theoretical prediction in 
\cite{Growth2, Growth11} that $\beta \in [0,1/2]$ has been verified for several economic systems, with empirical $\beta$
values typically in the range $0.1 < \beta < 0.3$ \cite{Growth11}.

\begin{figure}
\centering{\includegraphics[width=0.35\textwidth]{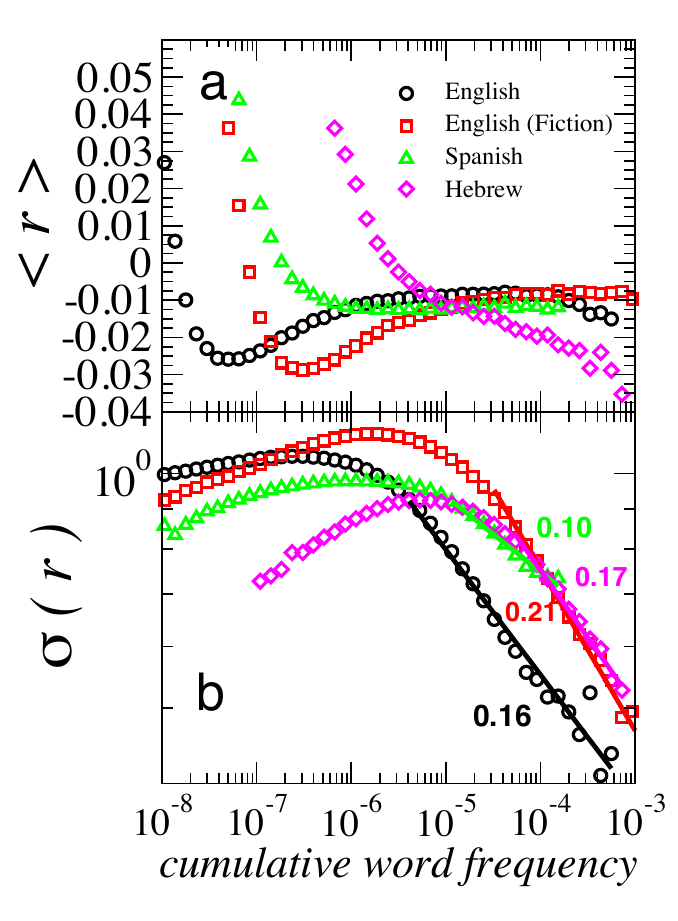}} 
\caption{ {\bf Scaling in the growth rate fluctuations of  words.} We show the dependence of growth rates on the cumulative word frequency $S_{i} \equiv \sum_{t'=0}^{t} f_{i}(t)$ using words satisfy the criteria $T_{i} \geq 10$ years. We
verify similar results for threshold values $T_{c} = 50, 100$, and $200$ years. {\bf (a)} Average growth rate $\langle r
\rangle$ saturates at relatively constant values for large $S$. {\bf (b)} Scaling in the standard deviation of growth rates $\sigma(r|S) \sim  S ^{-\beta}$ for words with large $S$. This scaling relation
is also observed for the growth rates of large economic institutions, ranging in size from companies to
 entire countries \cite{GrowthEcon,Growth11}. Here this size-variance relation corresponds to scaling exponent values $0.10 < \beta < 0.21$, which are related to the non-trivial bursting  patterns and non-trivial correlation patterns in literature topicality as indicated by the quantitative relation to the Hurst exponent, $H = 1-\beta$ shown in \cite{scalinghumaninteraction}. We calculate 
$\beta_{Eng.} \approx 0.16 \pm 0.01$, $\beta_{Eng. fict} \approx 0.21 \pm 0.01$,  $\beta_{Spa.} \approx 0.10 \pm 0.01$ and $\beta_{Heb.} \approx 0.17 \pm 0.01$. } 
\label{Rscaling}
\end{figure}

Since different words have varying lifetime trajectories as well as varying relative utilities, we now quantify how the
standard deviation $\sigma(r |S_{i})$ of  growth 
rates $r$ depends on the cumulative word frequency 
\begin{equation}
S_{i} \equiv \sum_{\tau=1}^{T_{i}} f_{i}(\tau) \ ,
\end{equation} 
of each word.
We choose this definition for proxy of ``word size'' since a writer can learn and recall a given word  through any of its historical uses.
Hence, $S_{i}$ is also proportional to the number of books in which word $i$ appears. This is significantly different than the assumptions of 
 replication null models (e.g. the Moran process) which use the concurrent frequency $f_{i}(t)$ as the sole factor determining the likelihood of future replication \cite{EvolLang,LanguageNeutralEvol}.

We estimate Eq. (\ref{sigmaRS})    by grouping words according to $S_{i}$ and then calculating the growth rate standard deviation  $\sigma(r | S_{i})$  for each group. 
Fig. \ref{Rscaling}(b) shows  scaling behavior consistent with Eq. \ref{sigmaRS} for large $S_{i}$, with $\beta \approx$ 0.10 -- 0.21 depending on
the corpus. A positive $\beta$ value means that words with larger cumulative word frequency have smaller annual growth rate fluctuations. 
 We conjecture that this statistical pattern emerges from 
the hierarchical organization of written language  \cite{WordnetlexiconHierarchy,SemanticNetwork,textHeirarchyCorr,FractalCorrCorpora,ReturnIntervalsLanguage} and the social properties of the speakers who use the words \cite{MetaHeaps,WordNiche,WordBursts}. 
As such, we calculate $\beta$ values that are consistent with
nontrivial correlations in word use, likely  related to the basic fact that books are 
topical  \cite{PlosOneZipf} and that book topics are correlated with cultural trends.\\

\noindent{\bf Quantifying the long-term cultural memory.} Recent theoretical work \cite{scalinghumaninteraction} shows that there is a fundamental relation between the
size-variance exponent  $\beta$ and the Hurst exponent $H$ quantifying the  auto-correlations in a stochastic
time series.   The novel relation $H = 1-\beta$ indicates that the temporal long-term persistence is intrinsically related
to the capability of the underlying mechanism  to absorb stochastic shocks.
Hence, positive  correlations ($H>1/2$) are predicted for non-trivial $\beta$ values (i.e. $0 \leq \beta \leq 0.5$). Note that
the Gibrat proportional growth model predicts $\beta = 0$ and that
a Yule-Simon urn model predicts $\beta = 0.5$ \cite{Growth11}.
Thus, $f_{i}(\tau)$ belonging to  words with large $S_{i}$
are predicted to show significant positive correlations, $H_{i}>1/2$.

To test this connection between  memory correlations and the size-variance scaling, we calculate the Hurst exponent $H_{i}$  for each time series belonging to the more relatively common words analyzed in
dataset (ii) using detrended fluctuation analysis (DFA)
\cite{DFA1,DFA2,scalinghumaninteraction}.   We plot in Fig. \ref{WordExample} the relative use time series $f_{i}(t)$ for the
words ``polyphony,'' ``Americanism,'' ``Repatriation,'' and ``Antibiotics''  along with DFA
curves  from which we calculate each $H_{i}$.   Fig. \ref{WordExample}(b) shows that the $H_{i}$ values for these four words are all
significantly greater than $H_{r} = 0.5$,  which is the expected Hurst exponent for a stochastic time series with no
temporal correlations. In Fig. \ref{DFAH}  we plot the distribution of $H_{i}$ values for the English fiction corpus and the
Spanish corpus. Our results are consistent with  the  theoretical prediction $\langle H \rangle = 1-\beta$ established in
\cite{scalinghumaninteraction} relating the variance of growth rates to the underlying temporal correlations in each
$f_{i}(t)$. Hence, we show that the  language evolution is fundamentally related to the complex features of cultural memory, i.e. the dynamics of cultural topic formation \cite{WordNiche,WordBursts,blogOmori,BlogWordDynamics} and  bursting \cite{Barabasibursts, SornettePNAS}.

\section*{Discussion}
With the digitization of written language, cultural trend analysis based around methods to extract quantitative patterns from  word counts is an emerging interdisciplinary field that has the  potential to provide novel insights into human sociology  \cite{PlosOneZipf,WordBursts,WordNiche,blogOmori,BlogWordDynamics,TwitterWords}. 
Nevertheless, the amount of metadata extractable from daily internet feeds is dizzying. This is highlighted by the  practical issue of defining objective significance levels to filter out the noise in the data deluge. For example, online blogs  can be vaguely categorized according to the coarse hierarchical schema: ``obscure blogs", ``more popular blogs", ``tech columns", and ``mainstream news coverage."   In contrast, there are well-defined entry requirements for published books and magazines, which must meet editorial standards and 
conform to the  principles of  market supply and demand. 
However, until recently, the vast  information captured in the annals of written language was largely inaccessible.

%Before the digitization of written language, the analysis of social and political trends required painstaking brute force manual work and rules of thumb so that a quantitative analysis of cultural trends would suffer from small sample effects since quantitative measurements would miss the large number of topics at any given point that are  below the threshold for detection, and long-term analysis would suffer from repeatedly crossings of topics above and below reliable significance thresholds due to intrinsic fluctuations.  
%However, the massive Google Books database allows social scientists to make reliable studies of word evolution at the ecosystem level which is far beyond the level of individual word case studies.
Despite the careful guard of libraries around the world, which house the written corpora for almost every written language, little is known about the aggregate dynamics of word evolution in written history.
Inspired by  research on the  growth patterns displayed by a wide range of competition driven systems -   from countries and  business firms \cite{GrowthEcon,Growth3,Growth1b,Growth6,Growth1,Growth2,Growth11,GrowthBuldPamm,Growth13b,lgcmps99} to religious activities \cite{Growth10}, universities \cite{Growth12}, scientific journals \cite{Growth7}, careers \cite{careerscaling} and  bird populations \cite{BirdGrowth} - here we extend the concepts and methods to word use dynamics. 

This study provides empirical evidence that words are competing actors in a system of finite resources.  Just as business firms compete for market share, words demonstrate the same growth statistics because they are competing for the use of the writer/speaker and for the attention of the corresponding reader/listener \cite{StatPhysLangDyn,ColorNaming,EmergenceTopicality,LanguageNeutralEvol,LangEcophysics}. 
A prime example of fitness-mediated evolutionary competition is the case of irregular and regular verb use in English.
By analyzing the regularization rate of irregular verbs through the history of the English language, Lieberman et al. 
 \cite{Evodynamicslanguage} show that the irregular verbs that are used more frequently are less likely to be overcome
  by  their regular verb counterparts. Specifically, they find that the  irregular verb death rate  scales as the inverse square root of the word's relative use.
A study of word  diffusion across Indo-European languages  shows  similar frequency-dependence of word replacement rates  \cite{lexdiffusion}.  

We document the case example of ÒX-ray,Ó  which shows how categorically related words can compete in a zero-sum game.  
Moreover, this competition does not occur in a vacuum. Instead, the dynamics are significantly related to diffusion and technology. 
Lexical diffusion occurs at many scales, both within relatively small groups and across nations \cite{lexdiffusion,WordNiche,LangEcophysics}.
The technological forces underlying word selection have changed significantly over the last 20 years.  With the advent of automatic spell-checkers 
in the digital era, words recognized by  spell-checkers  receive a significant boost in their ``reproductive fitness'' at the
 expense of their ``misspelled'' or unstandardized counterparts.  

We find that the dynamics  are influenced by historical context, trends in global communication, and the means for standardizing that communication.  Analogous to recessions and booms in a global economy, the marketplace for words waxes and wanes with a global pulse as historical events unfold.  And in analogy to financial regulations meant to limit risk and market domination, standardization technologies such as the dictionary and spell checkers serve as powerful arbiters in determining the characteristic properties of word evolution. Context matters, and so we anticipate that niches \cite{WordNiche} in various language ecosystems (ranging from spoken word to professionally published documents to various online forms such as chats, tweets and blogs) have heterogenous selection laws that may favor a given word in one arena but not another. Moreover, the birth and death rate of words and their close associates (misspellings, synonyms, abbreviations) depend on factors endogenous to the language domain such as correlations in word use to other partner words and polysemous contexts \cite{WordnetlexiconHierarchy, SemanticNetwork} as well as exogenous socio-technological factors and demographic aspects of the writers, such as age \cite{SemanticNetwork} and social niche \cite{WordNiche}.

We find a pronounced peak in the fluctuations of word growth rates when a word has reached approximately 30-50 years of age (see Fig. \ref{SDtrajectory}). We posit that this corresponds to the timescale for a word to be accepted into a standardized dictionary which inducts words that are used above a threshold frequency, consistent with the first-passage times to $f_c$ in Fig. \ref{SDtrajectory}(b). This is further corroborated  by the characteristic baseline frequencies associated with standardized dictionaries  \cite{googlepaper}. 
Another important timescale in evolutionary systems is the reproduction age of the interacting gene or meme host. Interestingly, a 30-50 year timescale is roughly equal to the characteristic human generational time scale.  The prominent role of new generation of speakers in language evolution has precedent in linguistics.  For example, it has been shown that primitive pidgin languages, which are little more than crude mixes of parent languages, spontaneously acquire the full range of complex syntax and grammar once they are learned by the children of a community as a native language.  It is at this point a pidgin becomes a creole, in a process referred to as nativization \cite{NowakLanguage}.

Nativization also had a prominent effect in the revival of the Hebrew language, a significant historical event which also manifests prominently in our statistical analysis.  The birth rate of new words in the Hebrew language jumped by a factor of 5 in just a few short years around 1920 following the Balfour Declaration of 1917 and the Second Aliyah immigration to Israel.  The combination of new Hebrew-speaking communities and political endorsement of a national homeland for the Jewish people in the Palestine Mandate had two resounding affects: (i) the Hebrew language, hitherto used largely only for (religious) writing, gained official status as a modern spoken language, and (ii) a centralized culture  emerged from this  national community.  The unique history of the Hebrew language in concert with the {\it Google Inc.} books data thus provide an unprecedented opportunity to quantitatively study the emerging dynamics of what is, in some regards, a new  language. 

The impact of historical context on language dynamics is not limited to emerging languages,  but extends to languages that have been active and evolving continuously for a thousand years.  We find that historical episodes can drastically perturb the properties of existing languages over large time scales. Moreover, recent studies  show  evidence for short-timescale cascading behavior in blog trends \cite{blogOmori,BlogWordDynamics}, analogous to the  aftershocks 
 following earthquakes and the cascades of market volatility  following financial news announcements \cite{OmoriFOMC}.  
The nontrivial autocorrelations and the leptokurtic growth distributions  demonstrate the significance of exogenous shocks which can result  in growth rates  that
significantly exceeding the frequencies that one would expect from non-interacting proportional growth models \cite{Growth1,Growth6}. 
 
A large number of the world's ethnic groups  are separated along linguistic lines.
 A language barrier can isolate its speakers by serving as a screen to  external events, which may further 
slow the rate of language evolution by  stalling endogenous change. Nevertheless, we find that the distribution of word growth rates significantly broadens during times of large scale conflict, revealed through the sudden increases in $\sigma(t)$ for the  English, French, German and Russian corpora during World War II \cite{LanguageAll}.  This can be understood as manifesting from the unification  of public consciousness that creates fertile breeding ground for new topics and ideas.  During war, people may be more likely to have their attention drawn to global issues.  Remarkably, the pronounced change during WWII was not observed for the Spanish corpus, documenting the relatively small roles that Spain and Latin American countries played in the war.

\section*{Methods}
\noindent{\bf Quantifying the word use trajectory.} 
Once a word is introduced into a language, what are the characteristic  growth patterns? 
To address this question, we first account for important variations in words, as the growth dynamics may depend on the frequency of the word as 
well as social and technological aspects of the time-period during which the word was born. 

 Here we define the age or trajectory year $\tau = t-t_{0,i}$ as the number of years after the word's first appearance in
the database. In order to compare  trajectories across time and across varying word frequency, we  normalize the
trajectories for each word $i$ by the average use
\begin{equation}
 \langle f_{i} \rangle \equiv \frac{1}{T_{i} } \sum_{t=t_{0,i}}^{t_{f,i}}f_{i}(t) 
\end{equation} 
over the lifetime $T_{i} \equiv t_{f,i}-t_{0,i}+1$ of the word, leading to the normalized trajectory,
\begin{equation}
f'_{i}(\tau) = f'_{i}( t - t_{i,0} | t_{i,0}, T_{i}) \equiv f_{i}( t - t_{i,0} )/\langle f_{i} \rangle \ .
 \label{fi}
\end{equation} 
By analogy, in order to compare various growth trajectories, we normalize the relative growth rate trajectory $r'_{i}(t)$ by the standard deviation
over the entire lifetime,
\begin{equation}
\sigma [r_{i}] \equiv \sqrt{\frac{1}{T_{i}} \sum_{t=t_{0,i}}^{t_{f,i}}[r_{i}(t)-\langle r_{i} \rangle]^{2} } \ .
\end{equation} 
Hence, the normalized relative growth trajectory is
\begin{equation}
r'_{i}(\tau) = r'_{i}( t - t_{i,0} | t_{i,0}, T_{i}) \equiv r_{i}( t - t_{i,0} )/\sigma [r_{i}]  \ .
 \label{ri}
\end{equation} 
 Figs. \ref{AveSDTraj}-\ref{AveSDTrajHeb} show the weighted averages $\langle f'(\tau | T_{c})\rangle$ and $\langle r'(\tau
| T_{c}) \rangle$ and the weighted standard deviations $\sigma[f'(\tau | T_{c})]$ and $\sigma[r'(\tau | T_{c})]$ calculated using normalized trajectories for new words in each corpus. We compute $\langle \dotsb \rangle $ and $\sigma[\dotsb]$  for each trajectory year  $\tau$ using all
$N_{t}$ 
trajectories  (Table \ref{TableSummary2}) that satisfy the criteria $T_{i} \geq T_{c}$
 and $t_{i,0} \geq 1800$. We compute the weighted average and the weighted standard deviation using
$\langle f_{i} \rangle$ as the weight value for word $i$, so that $\langle \dotsb \rangle $ and $\sigma[\dotsb]$ reflect the lifetime trajectories of the more common words that are ``new" to each corpus. 
 
Since there is an intrinsic word maturity $\sigma[r'(\tau | T_{c})]$ that is not accounted for in the quantity 
$r'_{i}(\tau)$, 
we further define the detrended relative growth
\begin{equation}
R \equiv r'_{i}(\tau) / \sigma[r'(\tau | T_{c})]
\label{R}
\end{equation}
which allows us to compare the growth factors for new words at various life stages. 
The result of this normalization is to rescale the standard deviations for a given trajectory year $\tau$ to unity for all 
values of  $r'_{i}(\tau)$. \\

\noindent{\bf Detrended fluctuation analysis of individual $f_{i}(t)$.}   Here we outline the DFA method for quantifying temporal 
autocorrelations in a general time series $f_{i}(t)$ that may have underlying trends, and compare the output with the  results expected from a 
 time series corresponding to a 1-dimensional random walk. 

 In a time interval 
$\delta t$, a time 
series  $Y(t)$ deviates from the previous value $Y(t-\delta t)$ by an amount $\delta Y(t) \equiv Y(t)-Y(t-\delta
t)$. A powerful result of the  
central limit theorem, equivalent to Fick's law of diffusion in 1 dimension, is that if the displacements are independent (uncorrelated corresponding to a simple Markov
process), then the total 
displacement $\Delta Y(t) = Y(t)-Y(0)$ from the initial location $Y(0) \equiv 0$ scales according to the total time $t$
as 
\begin{equation}
\Delta Y(t) \equiv Y(t)   \sim t^{1/2} \ .
\end{equation}
However, if there are  long-term correlations in the time series
$Y(t)$, then the relation is 
generalized to 
\begin{equation}
\Delta Y(t) \sim t^{H} \ ,
\end{equation}
 where $H$ is the Hurst exponent which corresponds to positive correlations for $H>1/2$
and negative 
correlations for $H<1/2$.

Since there may be underlying social, political, and technological trends that influence each time series $f_{i}(t)$, we use
the detrended fluctuation 
analysis (DFA) method \cite{DFA1, DFA2,scalinghumaninteraction} to analyze the residual fluctuations  $\Delta f_{i}(t)$ after we  
remove the local  trends. The method detrends the time series 
using time windows  of varying length   $\Delta t$. The time series $ \tilde{f}_{i}(t|\Delta t)$ corresponds to the
locally detrended time series using 
window size $\Delta t$. We  calculate the Hurst exponent $H$ using the relation
 between the root-mean-square displacement  $F(\Delta t)$ and the window size $\Delta t$  \cite{DFA1,
DFA2, scalinghumaninteraction},
\begin{equation}
F(\Delta t) = \sqrt{\langle \Delta \tilde{f}_{i}(t|\Delta t)^2\rangle} = \Delta t^H \ .
\label{Hurst}
\end{equation}
Here $\Delta \tilde{f}_{i}(t|\Delta t)$ is the local 
deviation 
 from the average trend, analogous to $\Delta Y(t)$ defined above. 
 
Fig. \ref{WordExample} shows 4 different $f_{i}(t)$  in panel (a), and plots the corresponding  $F_{i}(\Delta t)$
in 
 panel (b). The calculated $H_{i}$ values for these 4 words are all significantly
greater than the uncorrelated $H=0.5$ value, 
indicating strong 
 positive long-term correlations in the  use of these words, even after we have removed  the local trends using DFA. In these example cases, the
 trends are
related to political events 
such as war in the cases of ``Americanism'' and  ``Repatriation'', or the bursting associated with new technology in the
case of ``Antibiotics,'' or 
new musical trends illustrated in the case of ``polyphony.''

 In Fig. \ref{DFAH} we plot the pdf of $H_{i}$ values calculated for the
relatively common words 
analyzed in Fig. \ref{dRUPDF}(b). We also plot the pdf of  $H_{i}$ values calculated from shuffled time series, and these
values are centered around $ \langle H \rangle
\approx 0.5$ as expected from the removal of the intrinsic temporal ordering. Thus, using this method, we are able to
quantify the social memory 
characterized by the Hurst exponent which is related  to the bursting properties of linguistic trends,
and in general, to bursting phenomena  in human dynamics
\cite {blogOmori, BlogWordDynamics, Barabasibursts, SornettePNAS}.

\bigskip

\noindent {\bf Acknowledgments}\\
We thank Will Brockman, Fabio Pammolli, Massimo Riccaboni, and Paolo Sgrignoli for critical comments and insightful discussions. We gratefully acknowledge  financial support from the U.S. DTRA and the IMT Foundation. \\ %We also thank an anonymous referee at PNAS for extensive comments and critique.

\noindent {\bf Author Contributions}\\
\noindent  A. M. P., J. T., S. H. \& H. E. S., designed research, performed research, wrote, reviewed and approved the manuscript. A. M. P. and J. T. performed the numerical and statistical analysis of the data.\\

%\noindent {\bf Additional Information} \\

%\noindent {\bf Supplementary Information} Figs. S1-S7 and Tables S1-S3 accompany this paper.
%The authors declare no competing financial interests.

\clearpage
\newpage

\begin{widetext}

\pagebreak[4]

\clearpage
\newpage

\begin{center}

{\bf Supplementary Information} \\
\bigskip

{\bf \Large Statistical Laws Governing Fluctuations in Word Use \\ from Word Birth to Word Death} \\
\bigskip
Alexander M. Petersen,$^{1,2}$ J. Tenenbaum,$^{2}$ S. Havlin,$^{3}$ H. Eugene Stanley$^{2}$\\
$^{1}$IMT Lucca Institute for Advanced Studies, Lucca 55100, Italy \\
$^{2}$Center for Polymer Studies and Department of Physics, Boston University, Boston, Massachusetts 02215, USA\\
$^{3}$Minerva Center and Department of Physics, Bar-Ilan University, Ramat-Gan 52900, Israel\\
(2011)
\end{center}
\bigskip
\renewcommand{\theequation}{S\arabic{equation}
}
\renewcommand{\thefigure}{S\arabic{figure}}
\renewcommand{\thetable}{S\arabic{table}}
 
\setcounter{equation}{0}  % reset counter 
\setcounter{figure}{0}
\setcounter{table}{0}
\setcounter{section}{0}
\setcounter{page}{1}

\footnotetext[1]{ Corresponding author: Alexander M. Petersen \\
{\it E-mail}: \text{petersen.xander@gmail.com}
}

\begin{comment}
\section{Data Methods for analyzing new words}   
In order to analyze the lifetime trajectories of relatively new words, we select data according to several criteria
(i) We analyze words of length 20 or less in order to diminish the 
effects of insignificant strings of characters which can appear, e.g.
bababadalgharaghtakamminarronnkonnbronntonnerronntuonnthunntrovarrhounawnskawntoohoohoordenenthurnuk.  (ii) We only
consider words that have their first appearance $y_{0,i}$ after year $Y_{0} \equiv 1800$. We then compare words of varying
longevity, where the lifetime of word $i$ is $T_{i} \equiv t_{f,i}-t_{0,i}+1$, and group the words into 4   career
trajectory sets according to the four different thresholds  $T_{i} \geq T_{c}\equiv \{25, 50, 100, 200\}$ years. 
For the calculation of the average word-use trajectory $\langle u(\tau) \rangle$ and growth rate trajectory $\langle r'(\tau)
\rangle$, we use a sparsity threshold $s_{c} \equiv 0.2$ so that we consider words that have at most $s_{c} \cdot T_{i}$ years with no recorded use, 
corresponding to  $u_{i}(t)=0$ for year $t$. In the case that $u_{i}(t)=0$, we use the
approximation $f_{i}(t) \equiv f_{i}(t-1)$. Using these data cuts, we still have a significant number $N_{t}$ of unique
word trajectories to analyze, as shown in the data summary in  column 6 of Table \ref{TableSummary2}.
\end{comment}

\begin{figure}[h]
\centering{\includegraphics[width=0.7\textwidth]{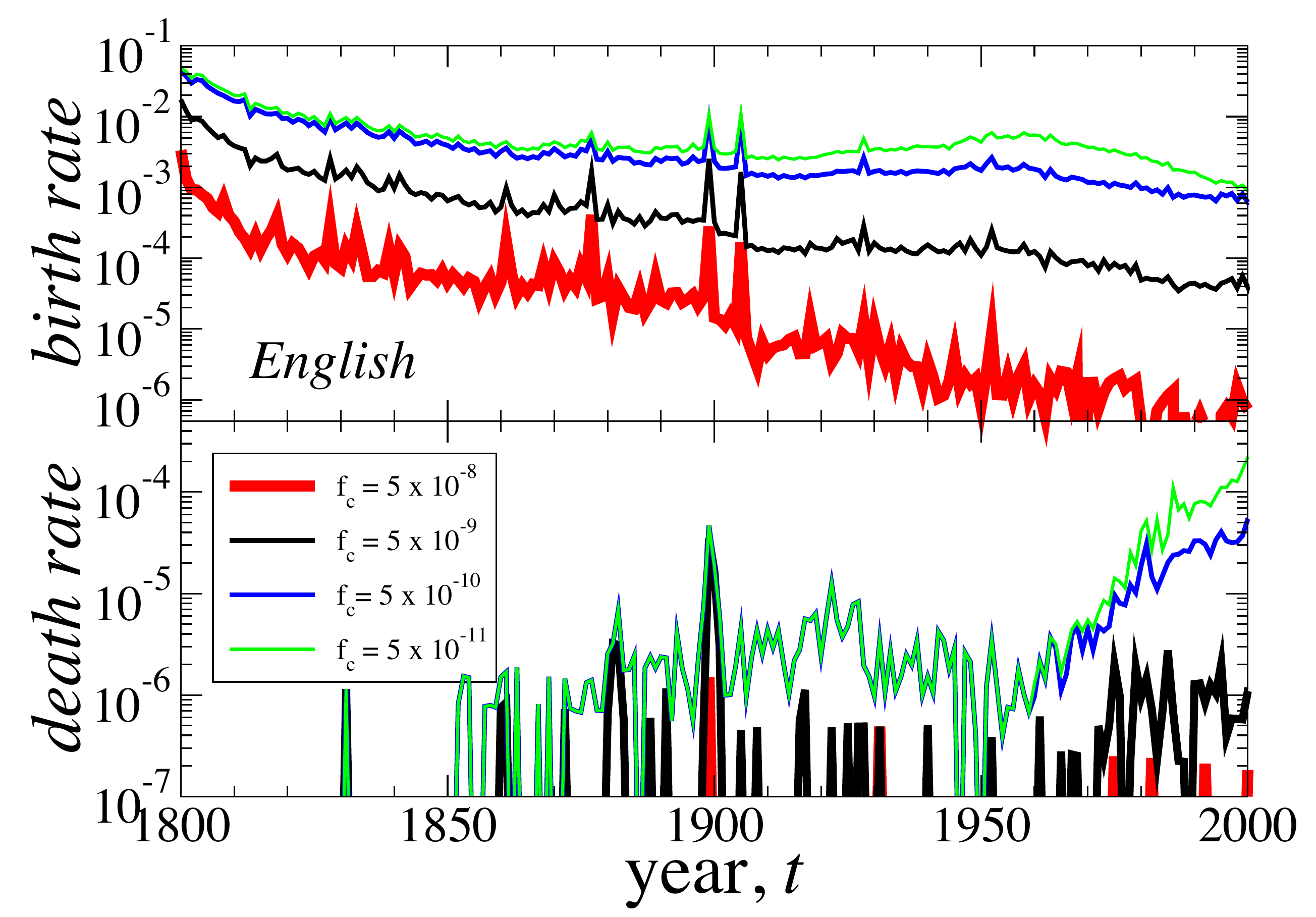}}
  \caption{ {\bf The birth and death rates of a word depends on the relative use of the word.} For the English corpus, we calculate the 
  birth and death rates for words with median lifetime relative use $\text{Med}(f_{i})$ satisfying $\text{Med}(f_{i}) > f_{c}$. The difference in the  birth rate curves corresponds to the contribution to the birth rate of words in between the two $f_{c}$ thresholds, and so
 the small difference in the curves for small $f_{c}$ indicates that the birth rate is largely comprised of words with relatively large $\text{Med}(f_{i})$. 
 Consistent with this finding, the largest contribution to the death rate is from words with relatively low $\text{Med}(f_{i})$.
 By visually inspecting the lists of dying words, we confirm that words with large relative use rarely become completely extinct (see Fig. \ref{radiology} for a counterexample word ``Roentgenogram'' which was once a frequently used word, but has since been eliminated due to
 competitive forces with other high-fitness competitors).} 
\label{BDRateFc}
\end{figure}

\begin{figure}
\centering{\includegraphics[width=0.7\textwidth]{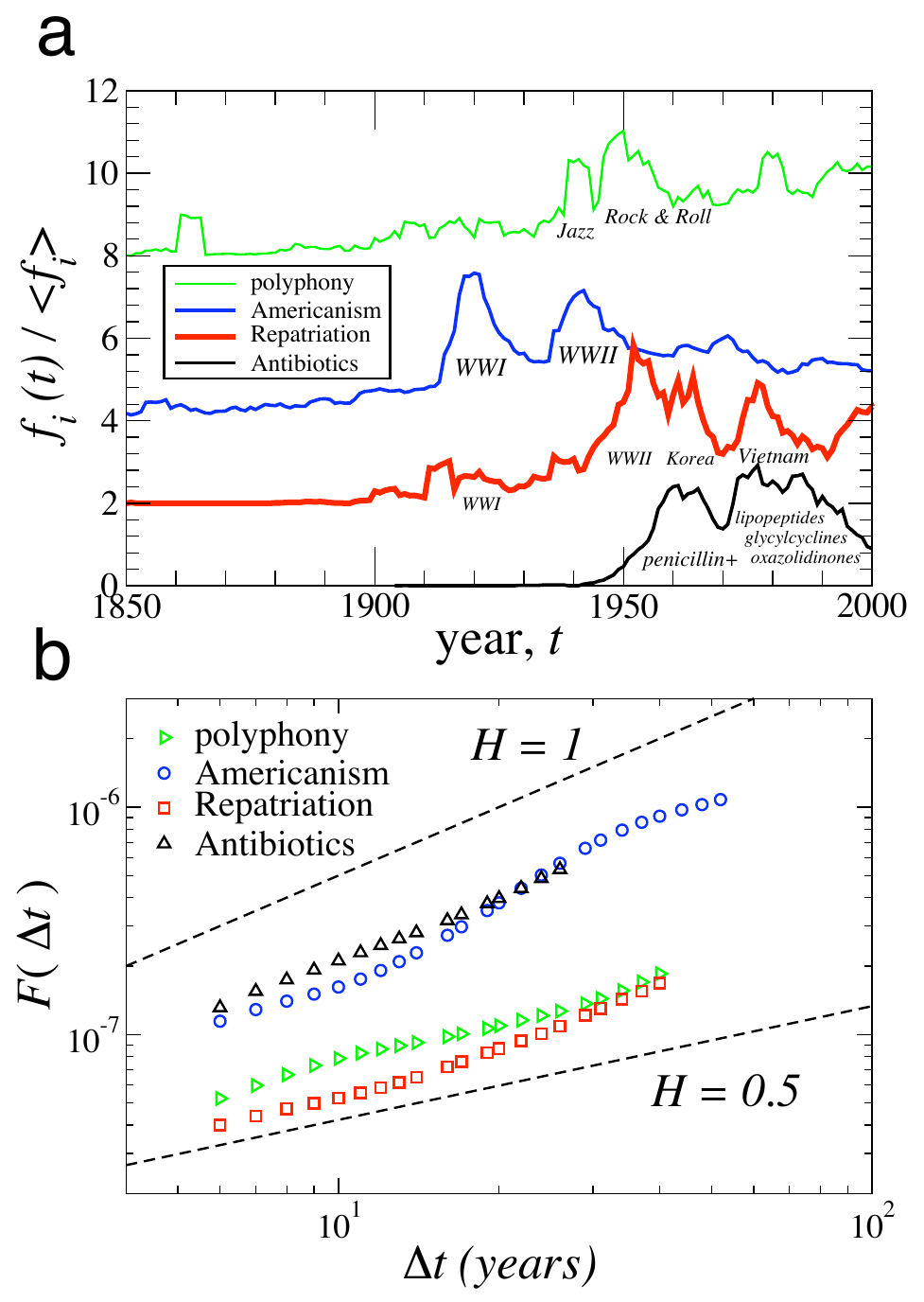}}
  \caption{ {\bf Measuring the social memory effect using the trajectories of single words.} We measure the Hurst exponent for individual $f_{i}(t)$ using the detrended fluctuation analysis method \cite{DFA1,DFA2,scalinghumaninteraction}. {\bf (a)} Four example $f_{i}(t)$, given in 
units of the average use $\langle f_{i} \rangle$, show bursting of use as a result of social and political ``shock'' events. 
We choose these four examples based on their relatively large $H_{i}>0.5$ values. The use of ``polyphony'' in the English corpus shows
peaks during the eras of  
jazz and rock and roll. The use of ``Americanism'' shows bursting during times of war, and the use
of ``Repatriation'' 
shows an approximate 10-year  lag in the bursting  after WWII and the Vietnam War. The use of the word
``Antibiotics''  is related to 
technological advancement. The top 3 curves are vertically displaced by a constant from the value $f_{i}(1800)  \approx
0$ so that the curves can be
distinguished.  {\bf (b)} We use detrended fluctuation analysis (DFA) to calculate the Hurst exponent $H_{i}$ for each word in order to quantify the long-term correlations (``memory'') in
each $f_{i}(t)$ time 
series. Fig. \ref{DFAH} shows the probability density function $P(H)$ of $H_{i}$ values
calculated for the 
 relatively common words found in English fiction and Spanish, summarized in Table \ref{TableSummary3}. } 
\label{WordExample}
\end{figure}

\begin{figure}
%\centering{\includegraphics[width=0.7\textwidth]{DFA_H_engfict_spa_PDF.pdf}}
\centering{\includegraphics[width=0.9\textwidth]{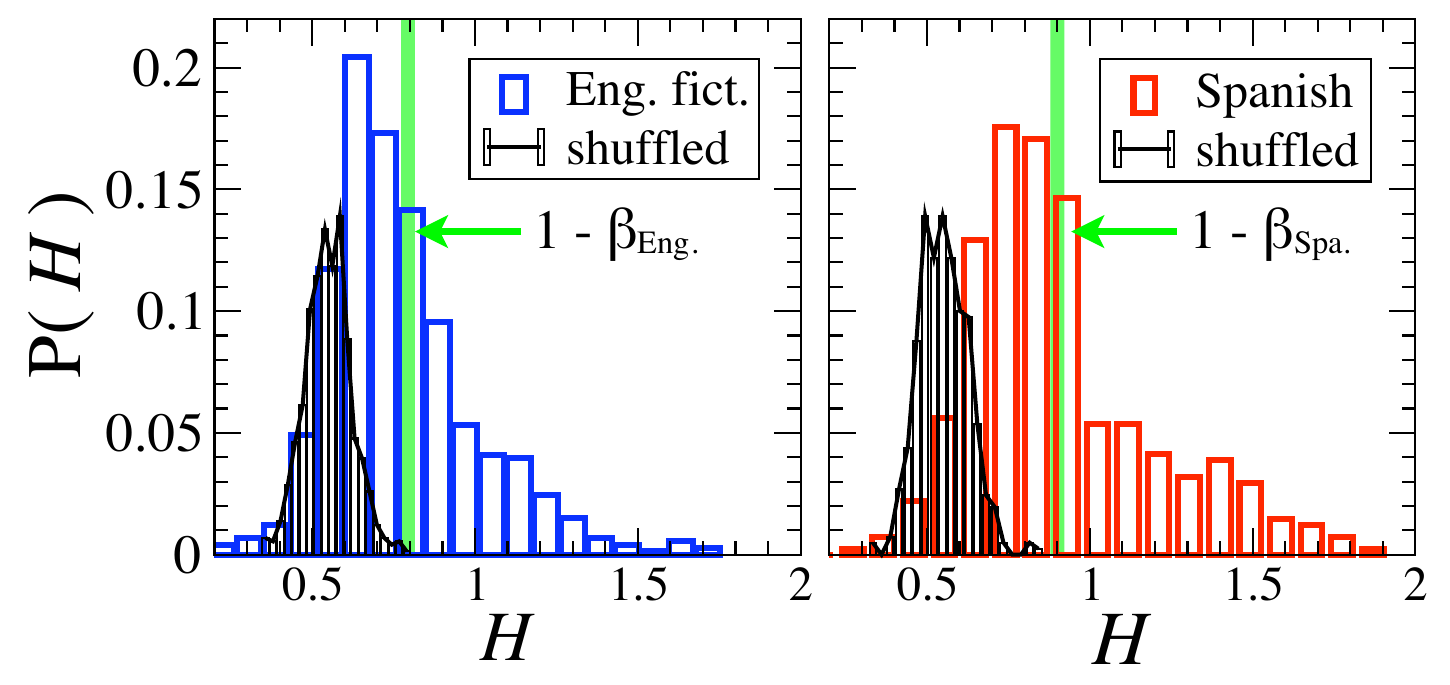}}
 \caption{\label{DFAH} {\bf Individual Hurst exponents $H_{i}$ indicate a strong positively correlated memory underlying word use dynamics}. Results of detrended fluctuation analysis (DFA) \cite{DFA1,DFA2, scalinghumaninteraction} on the common [dataset (ii)] words
analyzed in Fig. \ref {dRUPDF}(b) show 
 strong long-term memory with positive correlations, since $H > 1/2$, indicating strong correlated bursting in the
dynamics of word use, likely compounded by historical, social, or technological events. We calculate $\langle H_{i} \rangle \pm \sigma = 0.77 \pm 0.23$ (Eng.
fiction) and $\langle H_{i} \rangle = 0.90 \pm 0.29$ (Spanish).
The size-variance  $\beta$ values calculated from the data in Fig. \ref{Rscaling} confirm the theoretical prediction $\langle H \rangle = 1-\beta$ in  \cite{scalinghumaninteraction}. Fig. \ref{Rscaling} shows that
$\beta_{Eng. fict} \approx 0.21 \pm 0.01 $ and $\beta_{Spa.} \approx 0.10 \pm 0.01$.
  For the shuffled time series, we calculate  $\langle H_{i} \rangle \pm \sigma = 0.55 \pm 0.07$ (Eng. fiction) and
$\langle H_{i} \rangle \pm \sigma = 0.55 \pm 0.08$ (Spanish), which are consistent with time series that lack temporal
ordering (memory).
 % The top 2 panels correspond to $H$ values calculated using linear detrending, and the bottom 2 panels correspond to
%$H$ values calculated 
%using quadratic detrending.
}
\end{figure}

\begin{figure}
\centering{\includegraphics[width=0.49\textwidth]{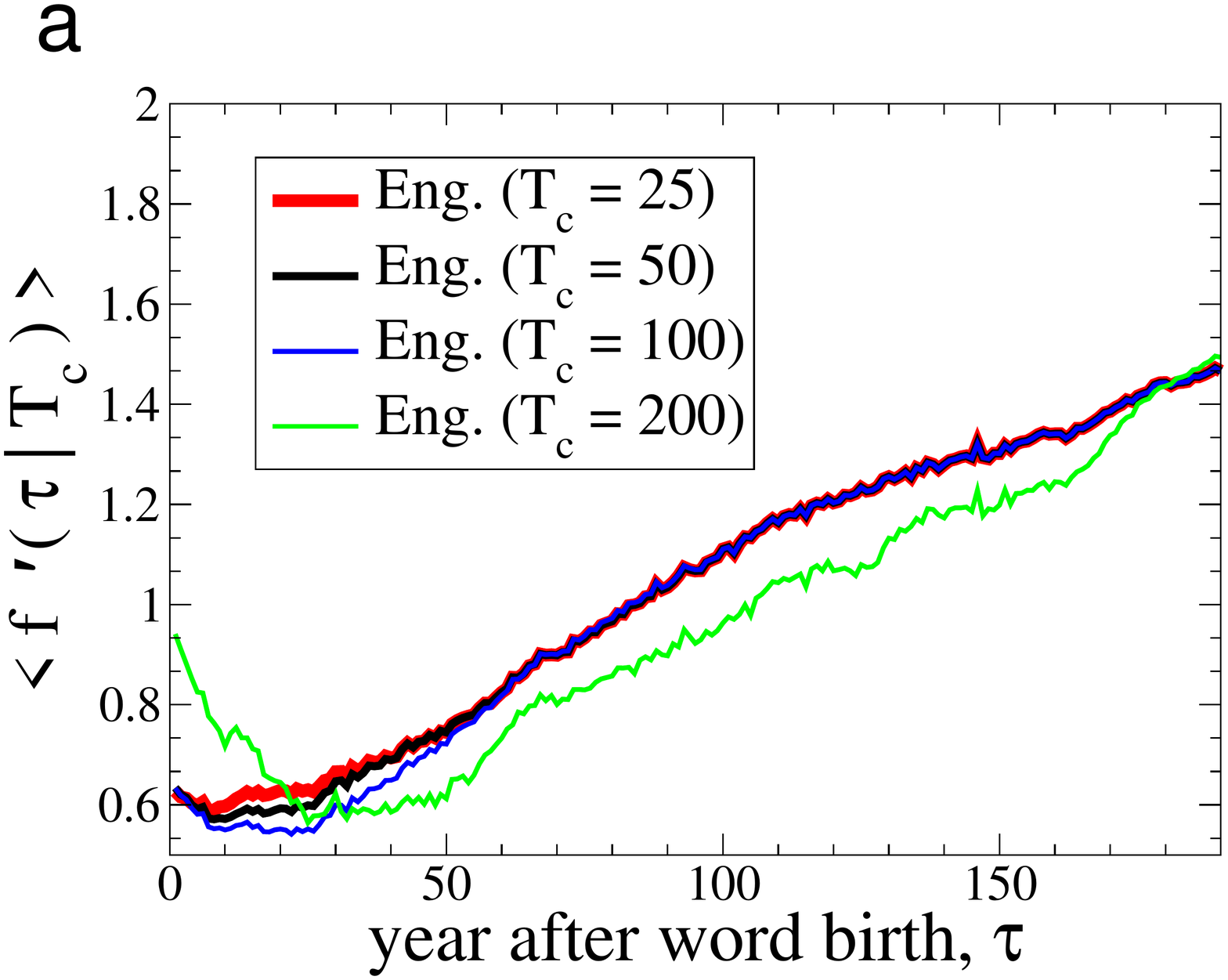}}
\centering{\includegraphics[width=0.49\textwidth]{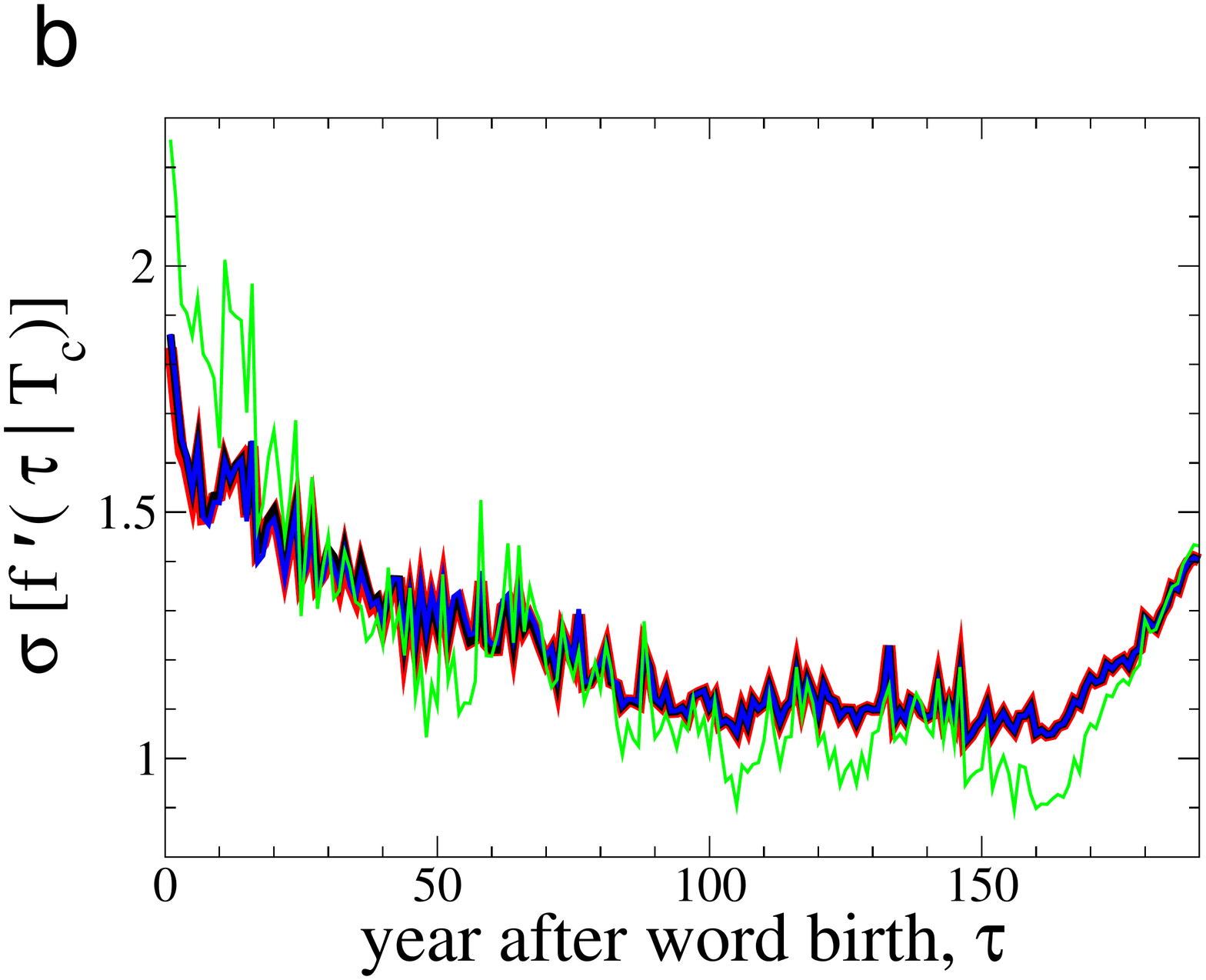}}

\centering{\includegraphics[width=0.49\textwidth]{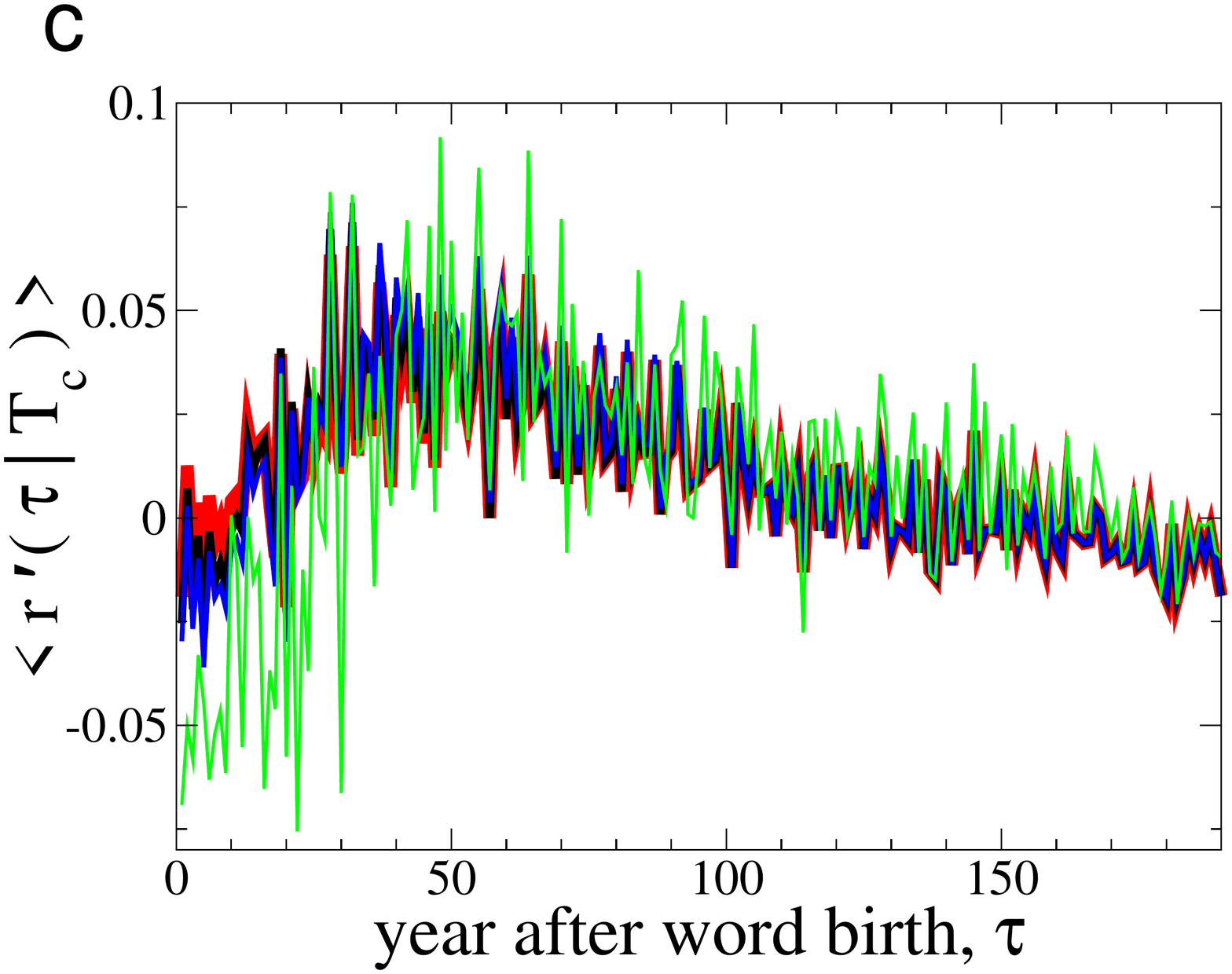}}
\centering{\includegraphics[width=0.49\textwidth]{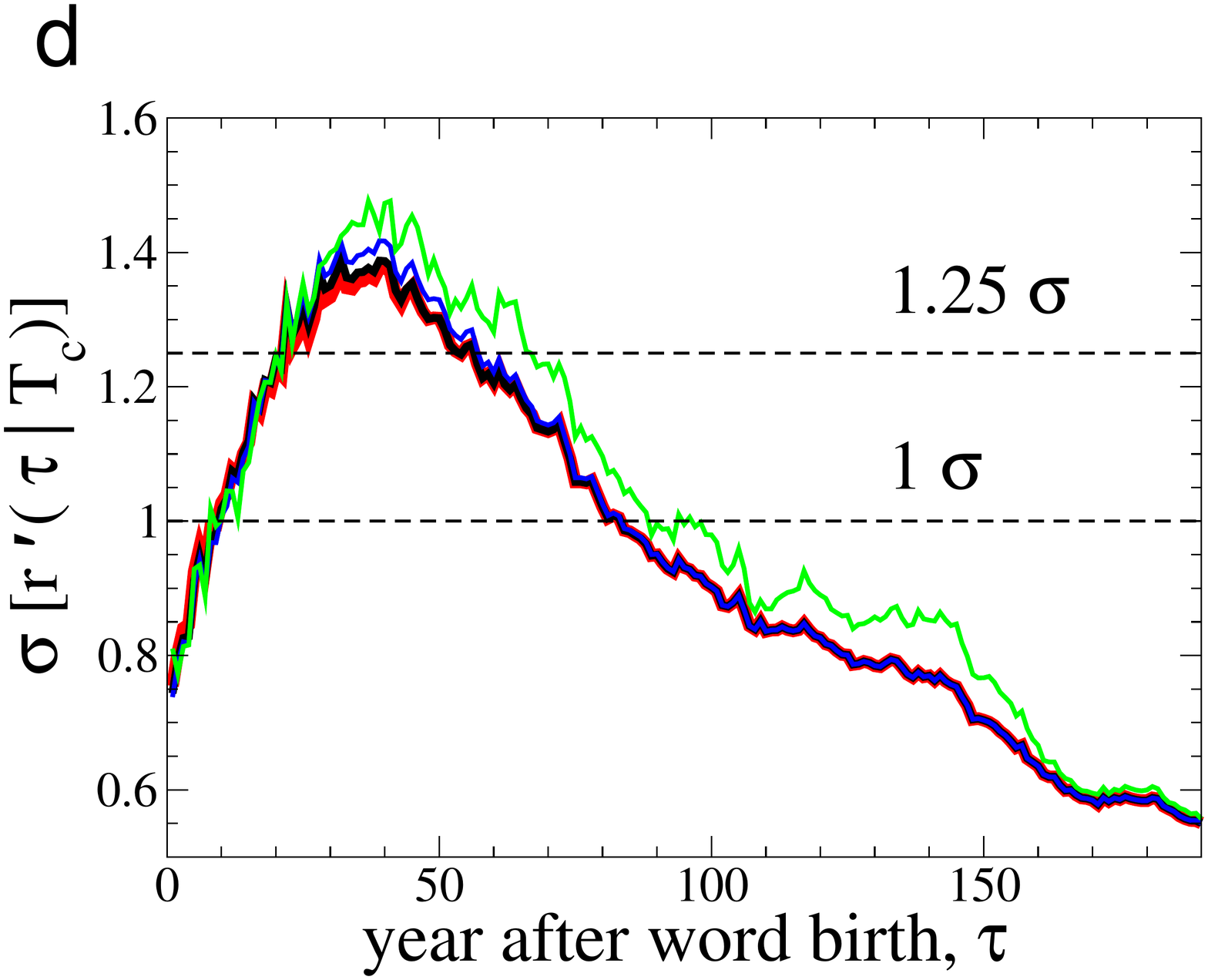}}
 \caption{ {\bf Statistical patterns in the growth trajectories of new words in the English corpus.} Characteristics of the time-dependent word trajectory show the time scales over which a typical word becomes
relevant or  fades. For  4 values of $T_{c}$, we show the word trajectories for dataset (i) words in the English corpus,
although the same qualitative
 results hold for the other  languages analyzed. Recall that $T_{c}$ refers to the subset of timeseries with lifetime
$T_{i} \geq T_{c}$, so that two trajectories calculated using different thresholds $T^{(1)}_{c}$ and $T^{(2)}_{c}$ only
vary for $\tau < Max[T^{(1)}_{c}, T^{(2)}_{c}]$. We show weighted average and standard deviations, using $\langle f_{i}
\rangle$ as the weight for word $i$ contributing to the calculation of each time series in year $\tau$. 
(a) The relative use increases with time, consistent with the definition of the weighted average which biases towards
words with large $\langle f_{i} \rangle$.  
For words with large $T_{i}$, the trajectory has a minimum which begins to reverse around $\tau \approx 40$ years,
possibly reflecting the amount of time it takes to 
reach a critical utility threshold that corresponds to a relatively high fitness value for the word in relation to its
competitors. (b) The variations in $\langle f(\tau | T_{c}) \rangle$ decrease with time reflecting the transition from the insecure
``infant'' phase to the  more secure ``adult'' phase in the lifetime trajectory. (c) The average growth trajectory is
qualitatively related to the logarithmic derivative of the curve in panel (a), and confirms that the region of largest
positive growth is $\tau \approx$ 30--50 years.  (d) The variations in the average trajectory are larger than 1.25
$\sigma$ for $30 \lesssim \tau \lesssim 50$ years and  are larger than 1.0 $\sigma$ for $10 \lesssim \tau \lesssim 80$
years. This regime of large fluctuations in the growth rates conceivably corresponds to the time period over which a
successful word is accepted into the standard lexicon, e.g. a word included in an official dictionary or an idea/event
recorded  in an encyclopedia or review.}
\label{AveSDTraj}
\end{figure}

\begin{figure}
\centering{\includegraphics[width=0.49\textwidth]{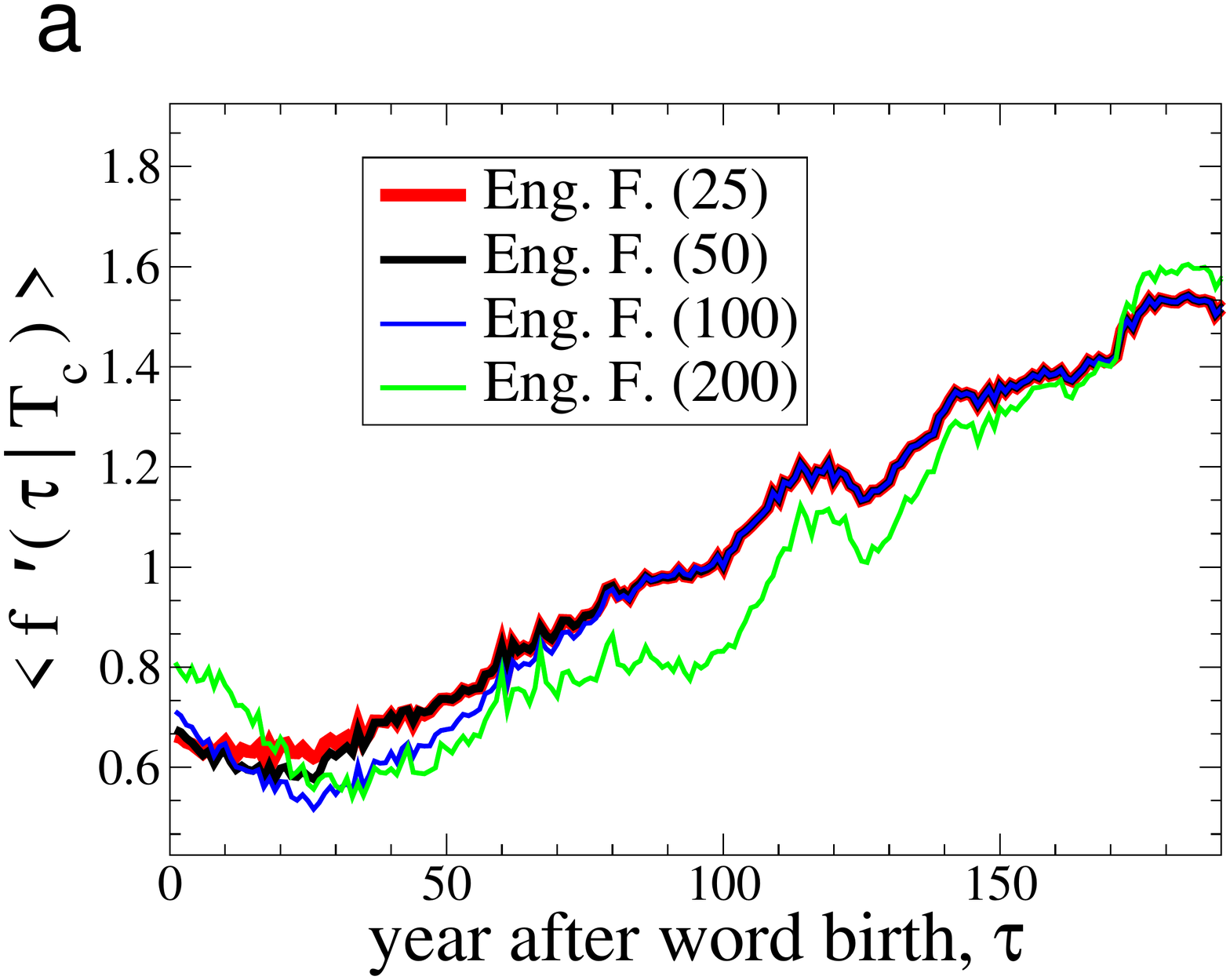}}
\centering{\includegraphics[width=0.49\textwidth]{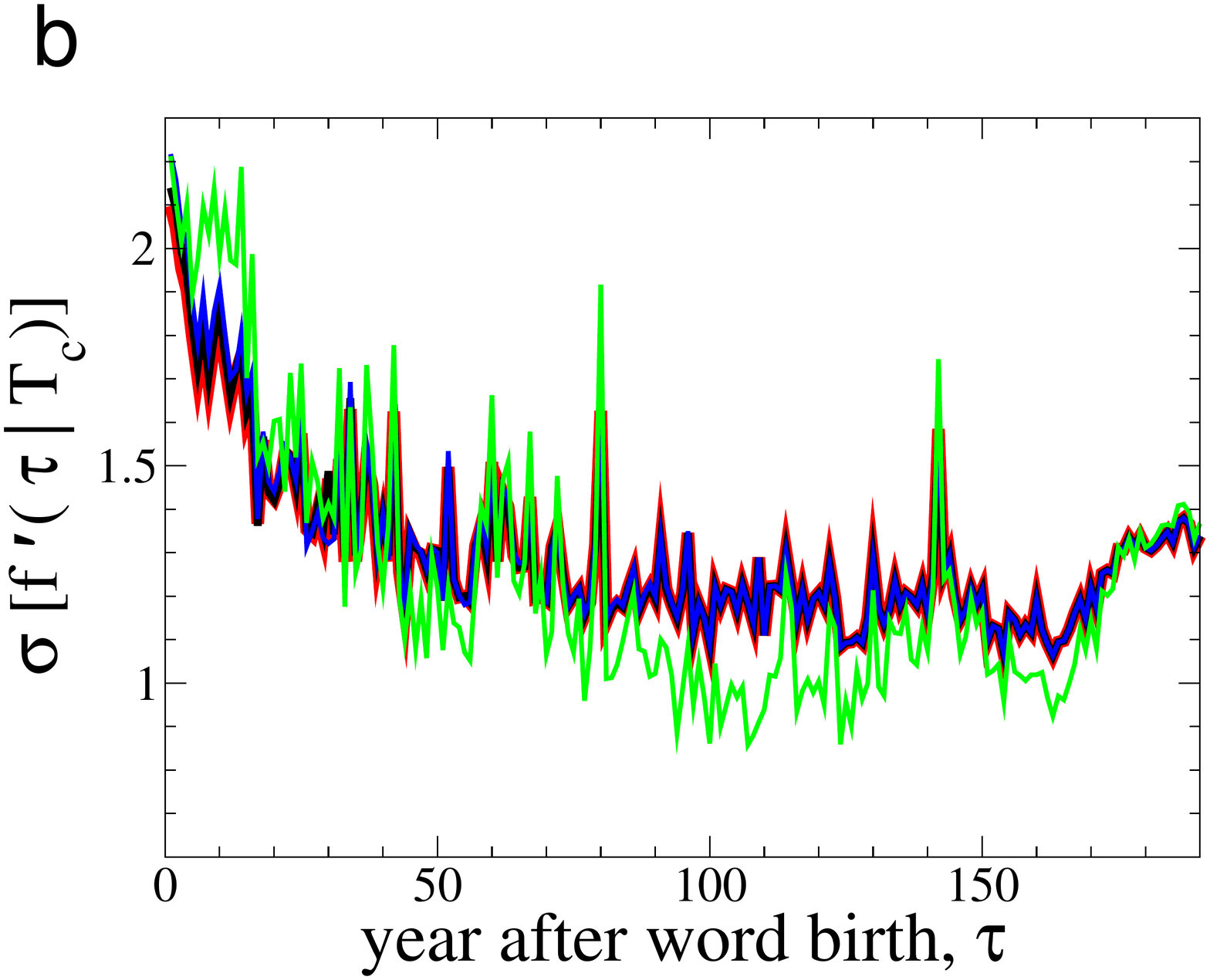}}

\centering{\includegraphics[width=0.49\textwidth]{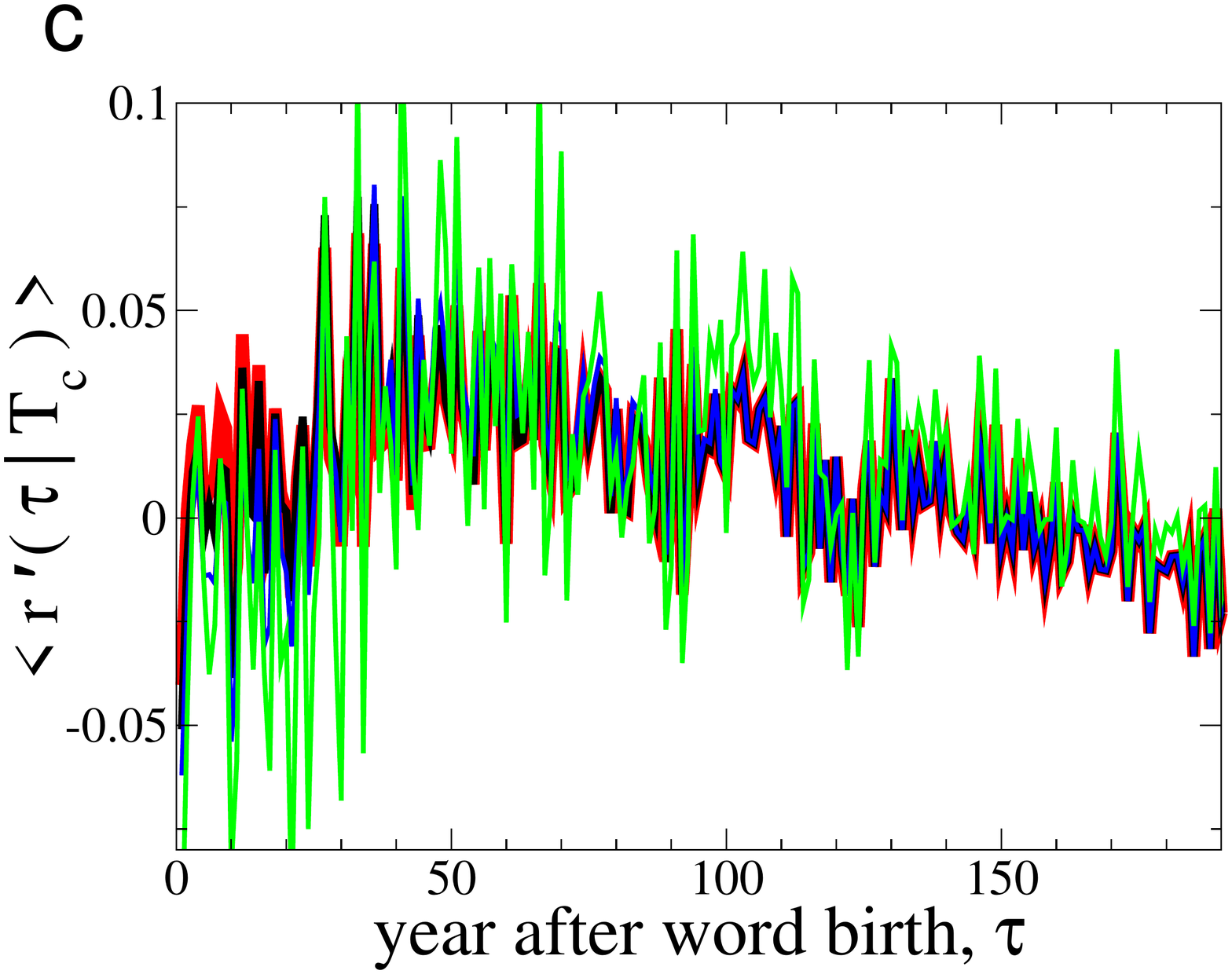}}
\centering{\includegraphics[width=0.49\textwidth]{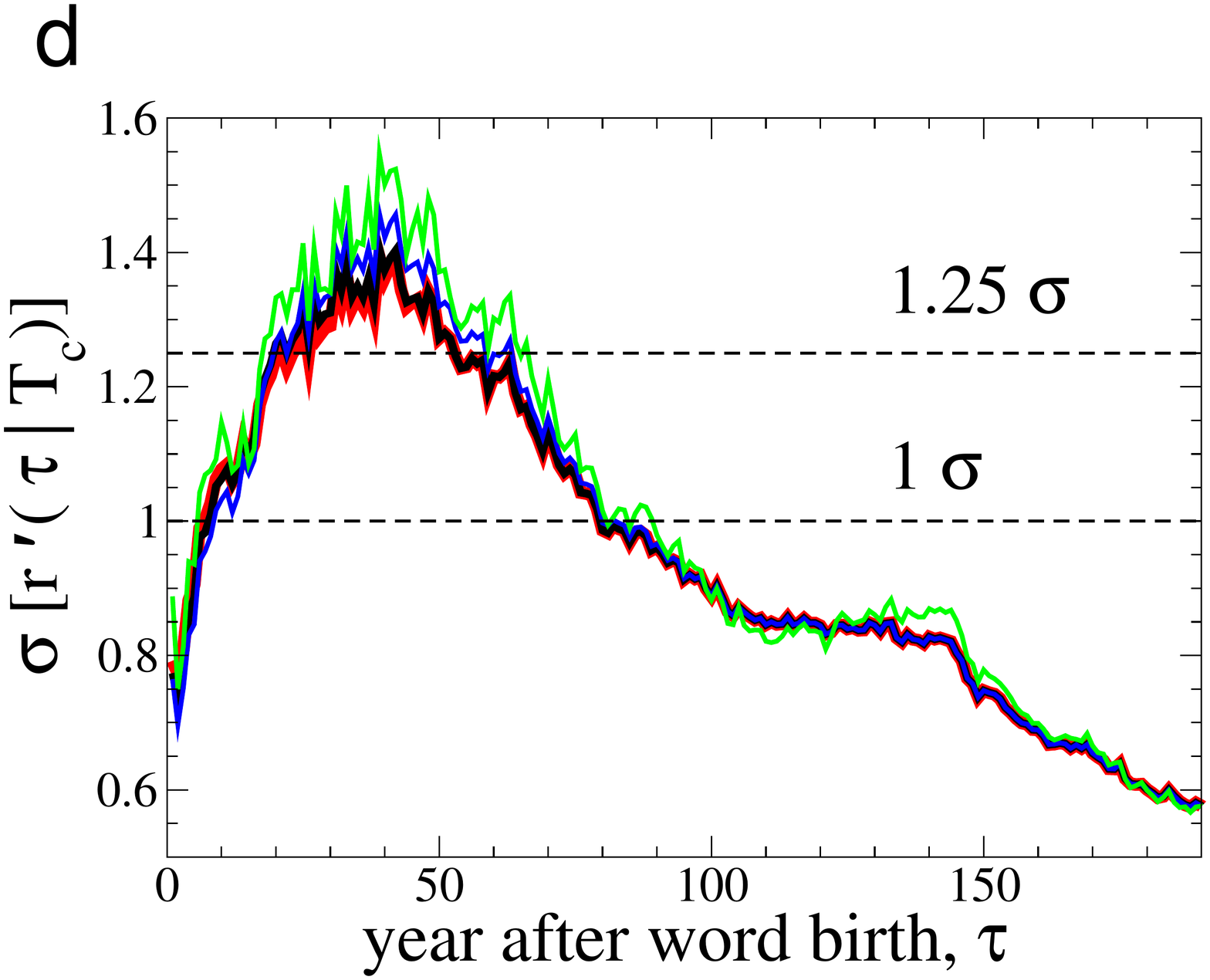}}
 \caption{ {\bf Statistical patterns in the growth trajectories of new words in the English Fiction corpus.} }
\label{AveSDTrajEngFict}
\end{figure}

\begin{figure}
\centering{\includegraphics[width=0.49\textwidth]{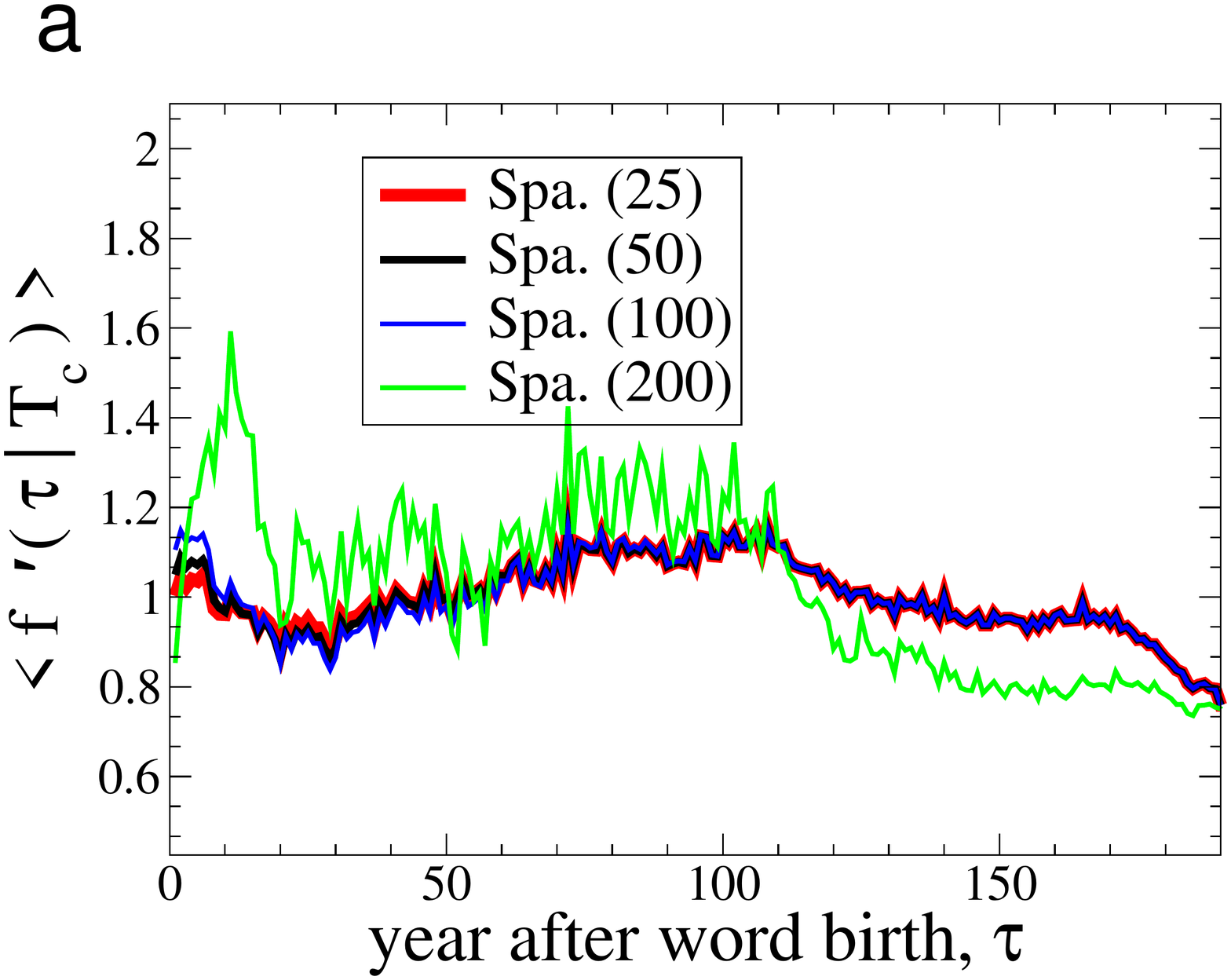}}
\centering{\includegraphics[width=0.49\textwidth]{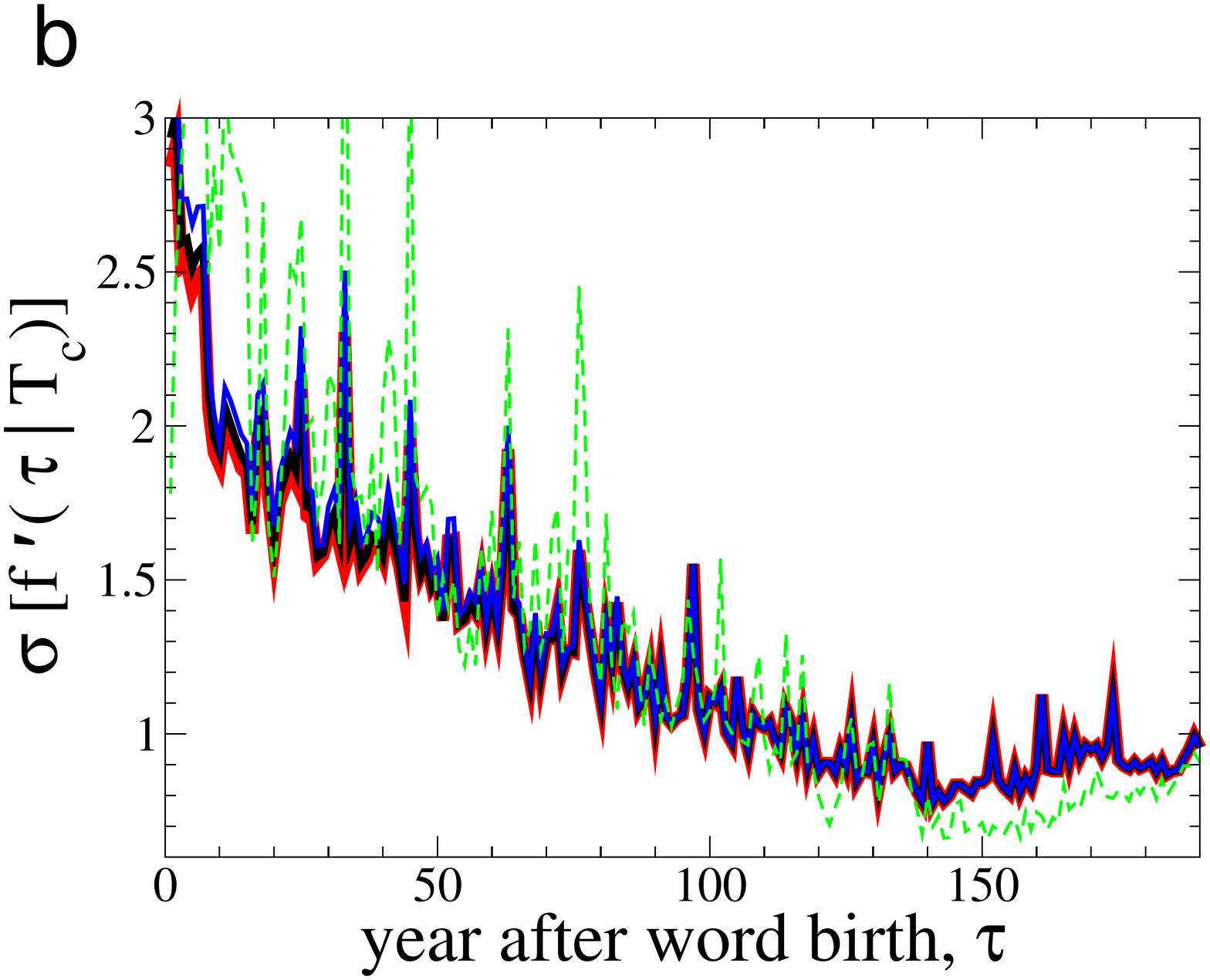}}

\centering{\includegraphics[width=0.49\textwidth]{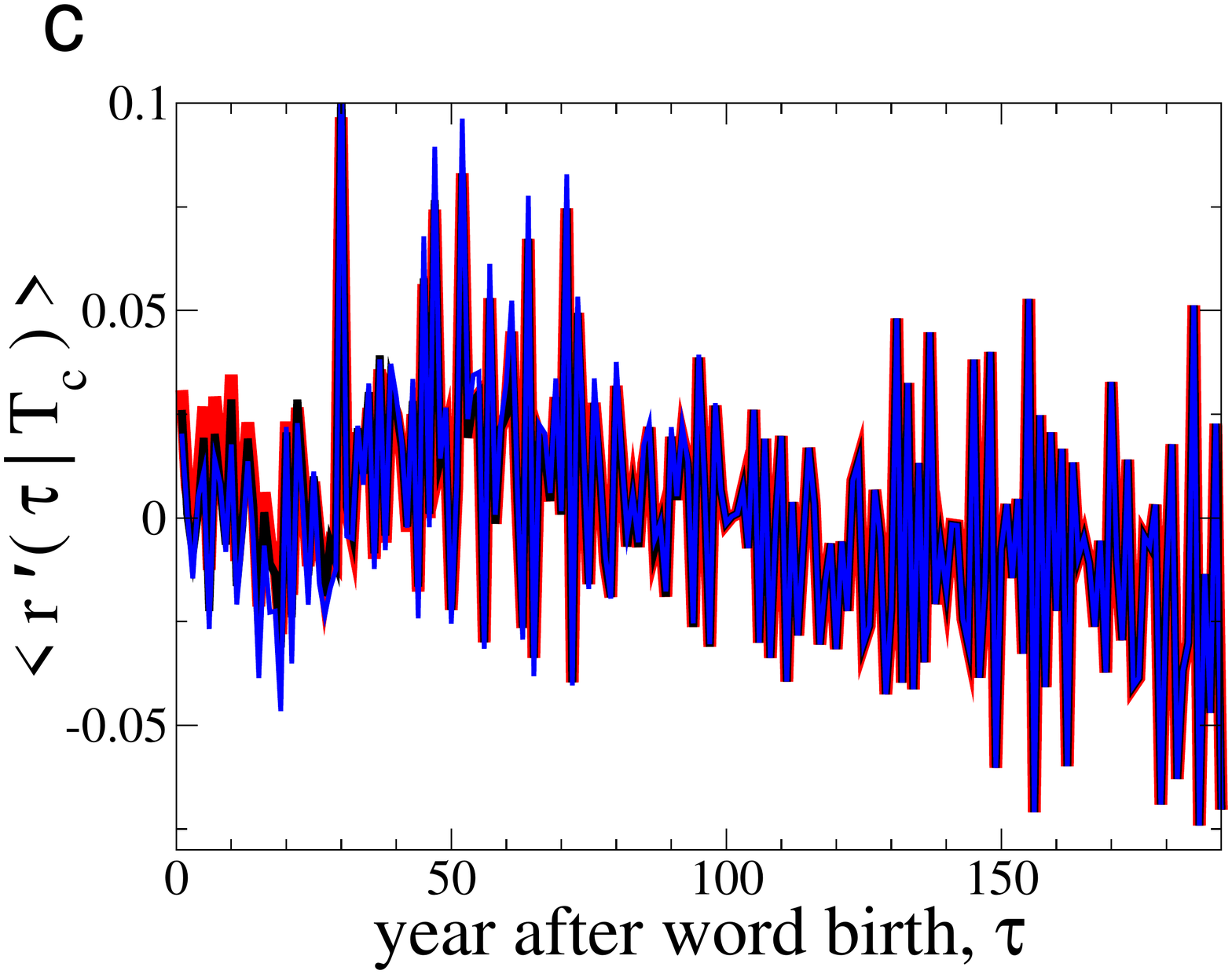}}
\centering{\includegraphics[width=0.49\textwidth]{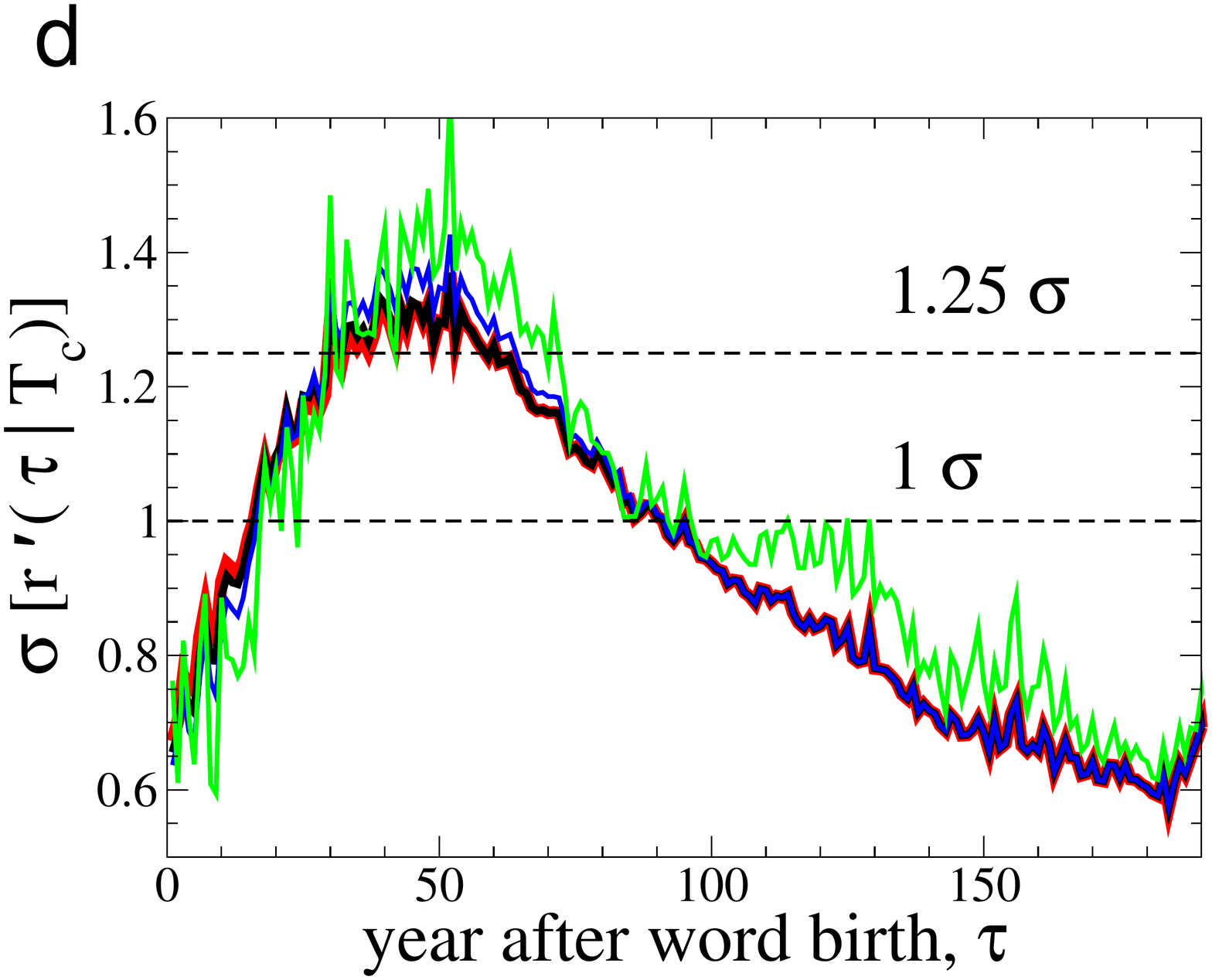}}
 \caption{ {\bf Statistical patterns in the growth trajectories of new words in the Spanish corpus.} }
\label{AveSDTrajSpa}
\end{figure}

\begin{figure}
\centering{\includegraphics[width=0.49\textwidth]{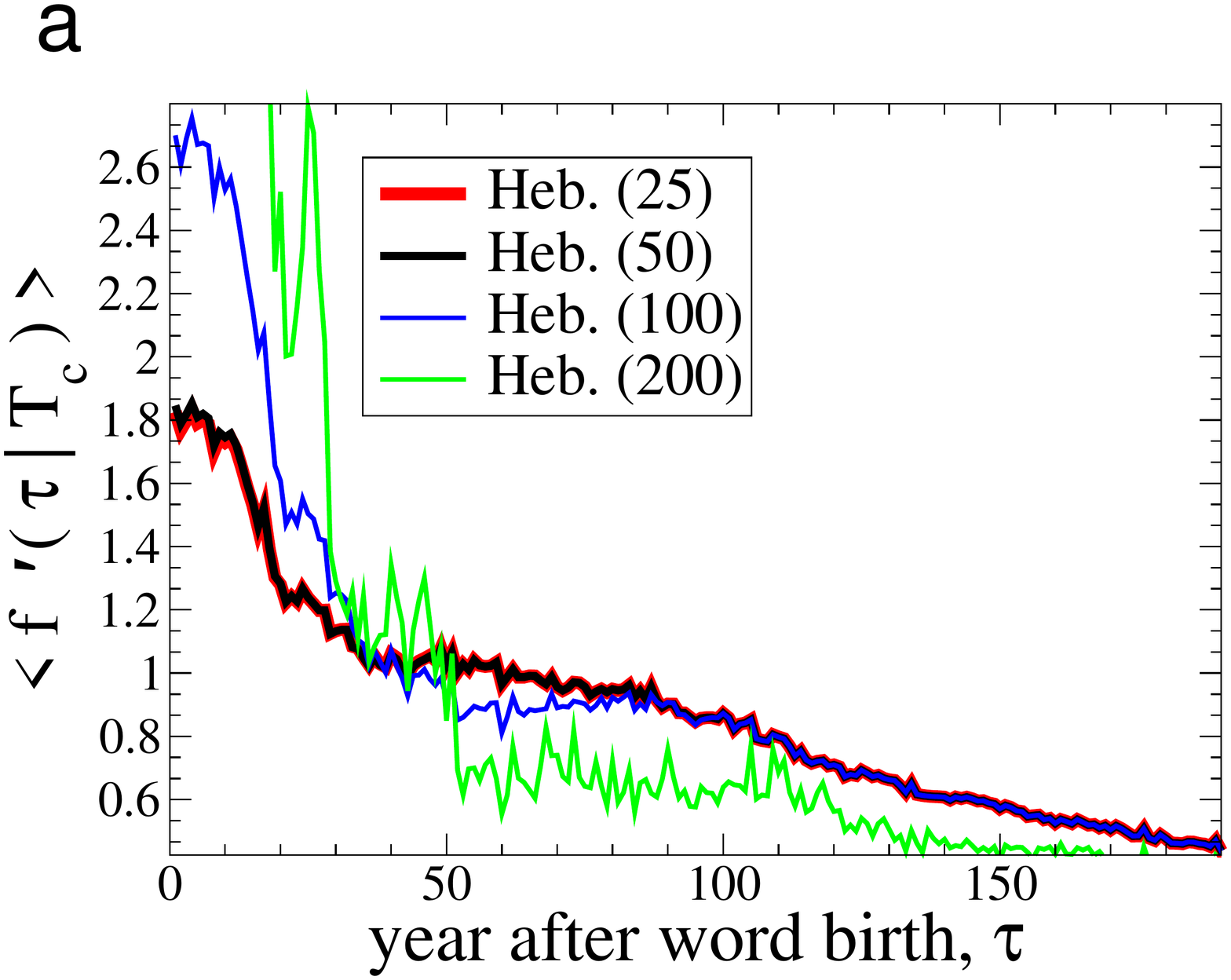}}
\centering{\includegraphics[width=0.49\textwidth]{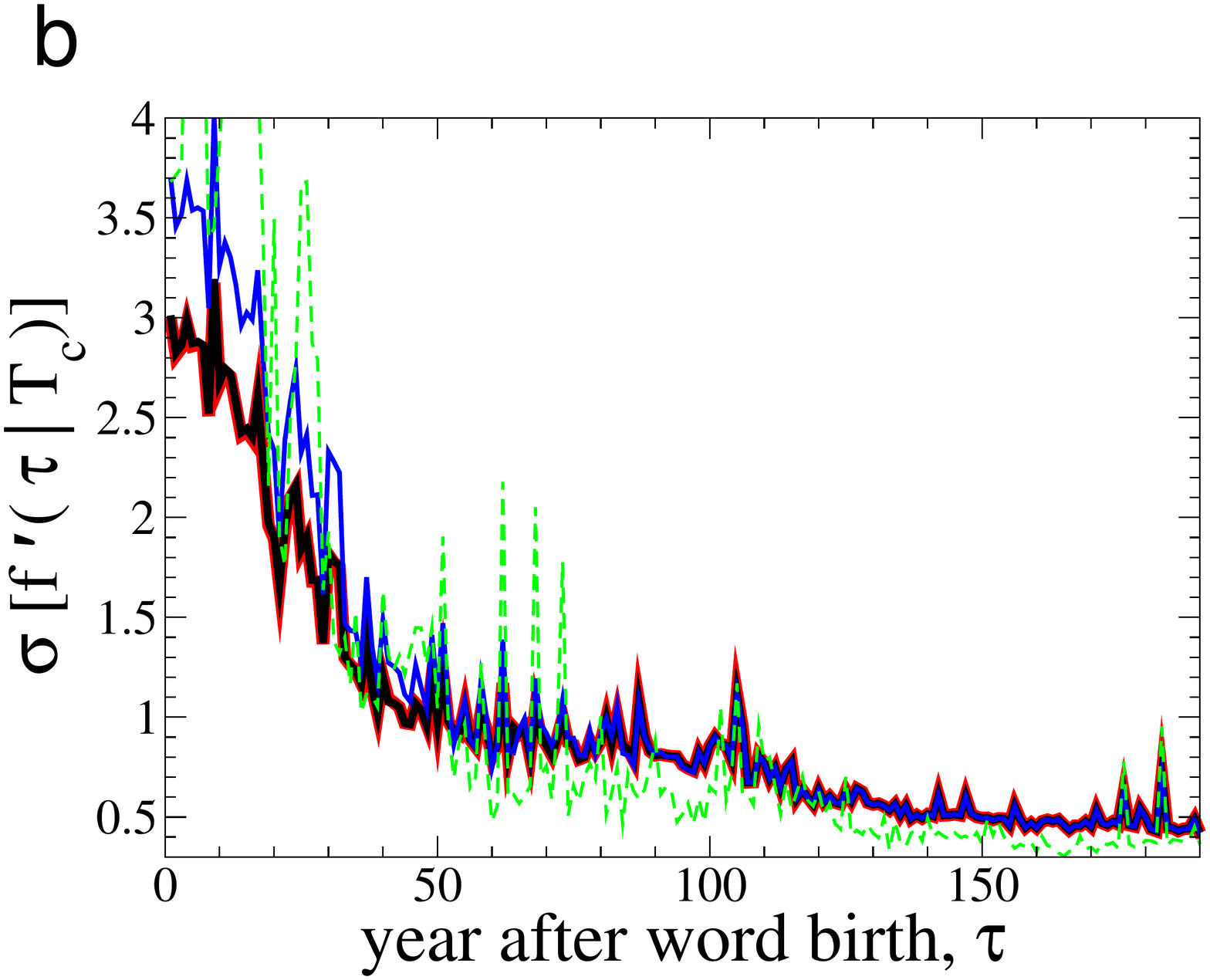}}

\centering{\includegraphics[width=0.49\textwidth]{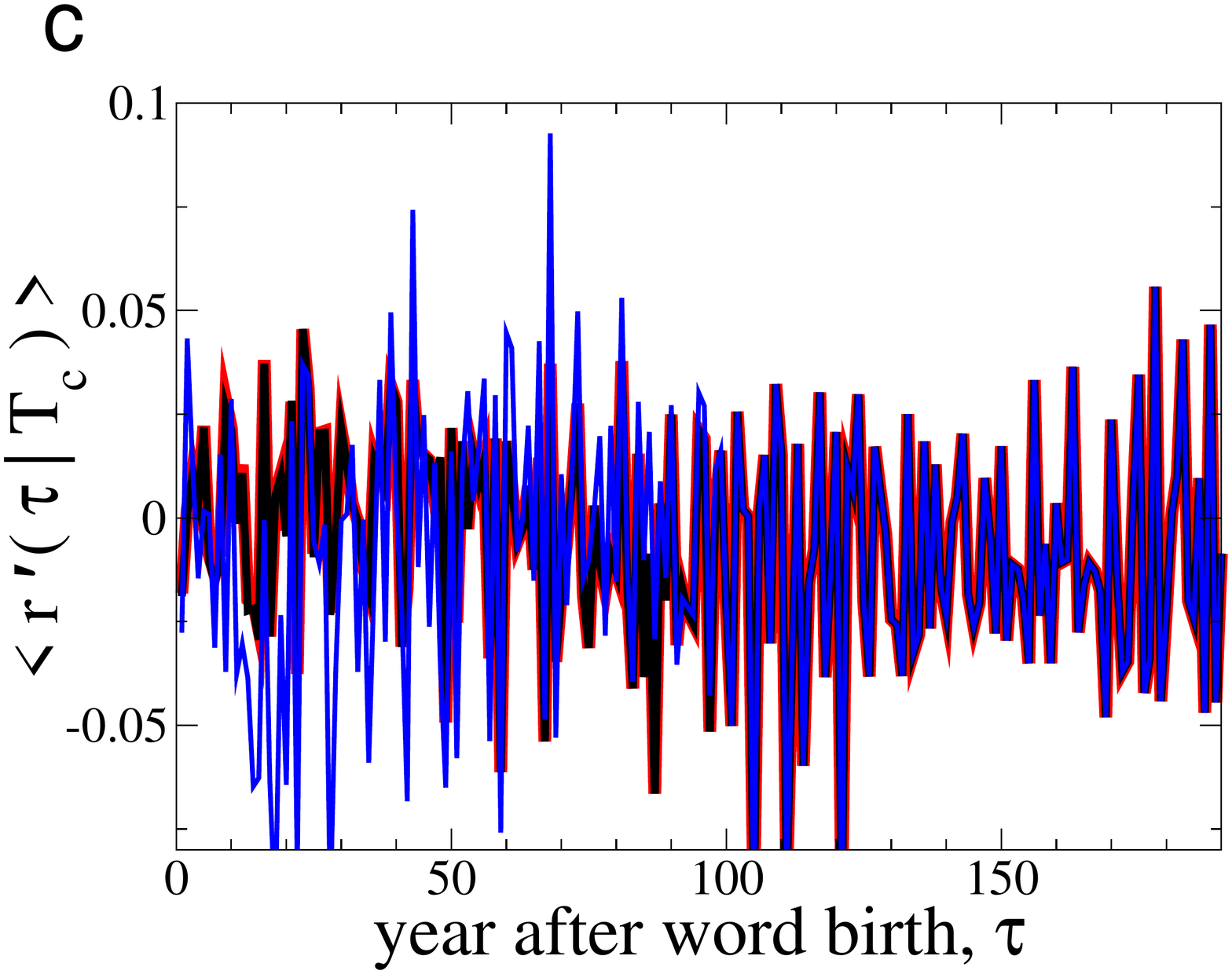}}
\centering{\includegraphics[width=0.49\textwidth]{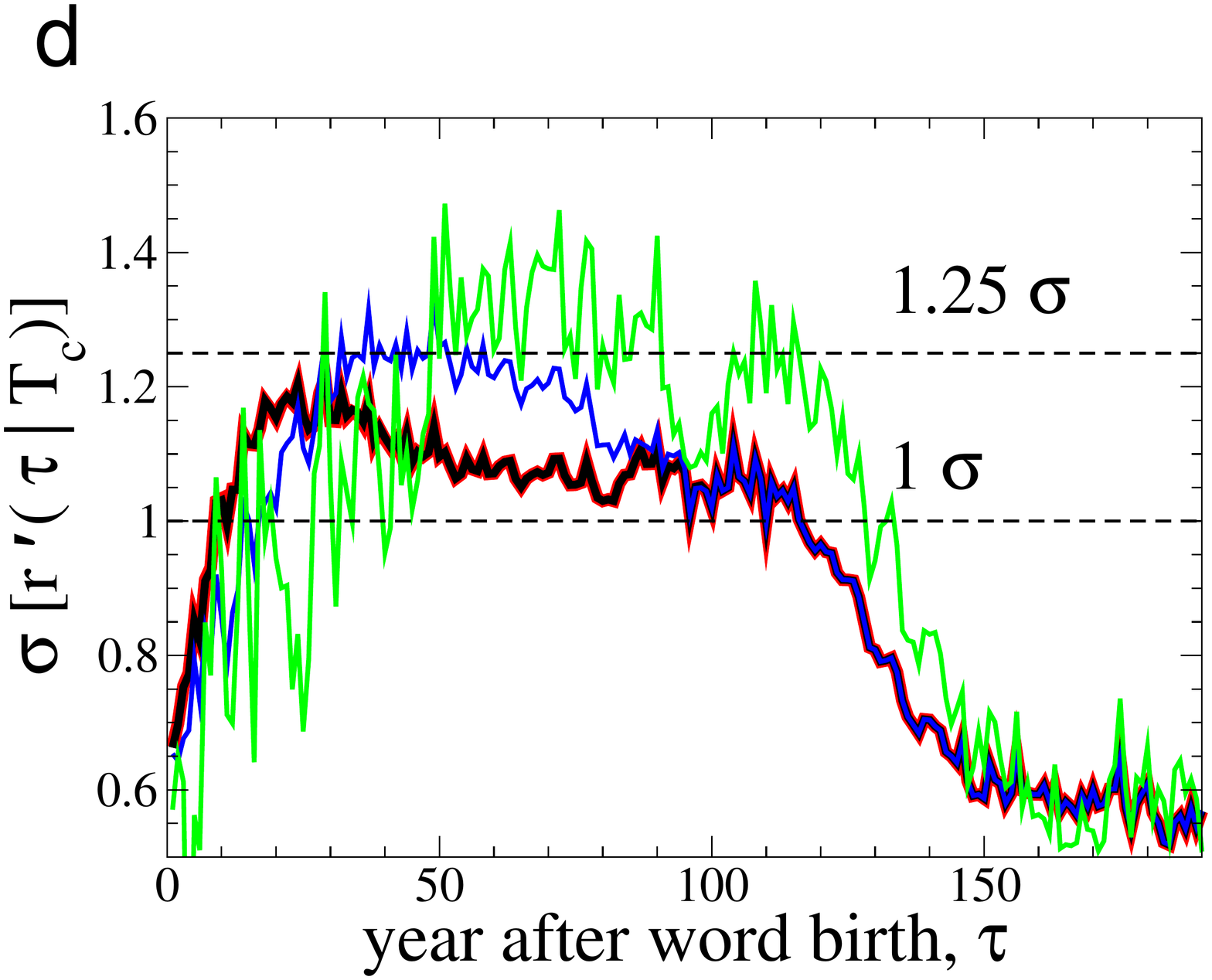}}
 \caption{ {\bf Statistical patterns in the growth trajectories of new words  in the Hebrew corpus.} }
\label{AveSDTrajHeb}
\end{figure}

%\begin{figure}
%\centering{\includegraphics[width=0.7\textwidth]{FirstPassageTau_Ave_tauc_100.pdf}}
%  \caption{ {\bf First passage time for new words.} The first passage time $\tau_{1}$ years for the relative use of a new word $i$ to exceed a given $f$-value, 
%defined as the first instance that satisfies $f_{i}[\tau_{1}(f)] \geq  f$, can also be used to 
 % quantify the thresholds for sustainability for new words.   The average first-passage time $\langle
%\tau_{1}(f) \rangle$ to $f_{c} \equiv 5 \times 10^{-8}$ for the English corpus, (recall  $f_{c}$ represents the threshold for a word belonging to the ``kernel'' lexicon), roughly corresponds to the peak time $\tau \approx 30-50$ years in $\sigma(\tau)$ shown in Fig. \ref{SDtrajectory}. This feature supports our conjecture that the peak in $\sigma(\tau)$ reflects the time scale over which a word is accepted into the standard lexicon.   } 
%\label{1passage}
%\end{figure}

\clearpage
\newpage

\begin{table}
\caption{  Summary of annual growth trajectory data for varying  threshold $T_{c}$, and  $s_{c}=0.2$, $Y_{0}\equiv 1800$
and $Y_{f}\equiv 2008$.}
\begin{tabular}{@{\vrule height 10.5pt depth4pt  width0pt}lc|c|c|c|c|c|} \multicolumn7c{Annual growth $R(t)$ data}\\
\noalign{
\vskip-11pt} Corpus,\\
\cline{2-7}
\vrule depth 6pt width 0pt (1-grams)& $T_{c} (years)$ & $N_{t}(words)$ & \% (of all words)  & $N_{R}(values)$ & $\langle
R \rangle$ & $\sigma[R]$ \\
\hline 
English  &          25 &   302,957 &  4.1 &  31,544,800 &  $2.4\times 10^{-3}$ &  1.00 \\
English fiction & 25 &   99,547   & 3.8  &  11,725,984 &  $-3.0\times 10^{-3}$ &  1.00 \\
Spanish &           25 &    48,473 &  2.2 &  4,442,073 &  $1.8\times 10^{-3}$ &  1.00 \\
Hebrew &           25 &    29,825 &  4.6 &  2,424,912 &  $-3.6\times 10^{-3}$ &  1.00 \\
\hline English  &          50 &   204,969 &  2.8 &  28,071,528 &  $-1.7\times 10^{-3}$ &  1.00 \\
English fiction & 50 &     72,888 &  2.8 &  10,802,289 &  $-1.7\times 10^{-3}$ &  1.00\\
Spanish &           50 &     33,236 &  1.5 &  3,892,745 &  $-9.3\times 10^{-4}$ &  1.00 \\
Hebrew &           50 &     27,918 &  4.3 &  2,347,839 &  $-5.2\times 10^{-3}$ &  1.00 \\
\hline English  &          100 &     141,073 &  1.9 &  23,928,600 &  $1.0\times 10^{-4}$ &  1.00 \\
English fiction & 100 &    53,847 &  2.1 &  9,535,037 &  $-8.5\times 10^{-4}$ &  1.00 \\
Spanish &           100 &    18,665 & 0.84  &  2,888,763 &  $-2.2\times 10^{-3}$ &  1.00 \\
Hebrew &           100 &    4,333 & 0.67  &  657,345 &  $-9.7\times 10^{-3}$ &  1.00 \\
\hline English  &          200 &     46,562 &  0.63 &  9,536,204 &  $-3.8\times 10^{-3}$ &  1.00 \\
English fiction & 200 &    21,322 &  0.82 &  4,365,194 &  $-3.5\times 10^{-3}$ &  1.00 \\
Spanish &           200 &    2,131 &  0.10 &  435,325 &  $-3.1\times 10^{-3}$ &  1.00 \\
Hebrew &           200 &    364 &  0.06 &  74,493 &  $-1.4\times 10^{-2}$ &  1.00 \\
\hline\\
\end{tabular}
\label{TableSummary2}
\end{table}

\begin{table}
\caption{  Summary of  data for the relatively common words that meet the criterion that their average word use $\langle
f_{i} \rangle$ over the 
entire word history is larger than a threshold $f_{c}$, defined for each corpus.   In order to select relatively frequently used
words, we use the following three 
criteria: the word lifetime $T_{i} \geq 10$ years, $1800 \leq t \leq 2008$, and $\langle f_{i} \rangle \geq f_{c}$. 
 }
\begin{tabular}{@{\vrule height 10.5pt depth4pt  width0pt}lc|c|c|c|c|c|} \multicolumn6r{ Data summary for relatively
common words}\\
\noalign{
\vskip-11pt} Corpus,\\
\cline{2-7}
\vrule depth 6pt width 0pt (1-grams)& $f_{c}$ & $N_{t}(words)$ & \% (of all words)  & $N_{r'}(values)$ & $\langle r'
\rangle$ & $\sigma[r']$ \\
\hline
\hline English & $5 \times 10^{-8}$ & 106,732 & 1.45 & 16,568,726 &  1.19 $\times 10^{-2}$ & 0.98 \\
English fiction & $1 \times 10^{-7}$ & 98,601 & 3.77 & 15,085,368 & 5.64 $\times 10^{-3}$ & 0.97 \\
Spanish & $1 \times 10^{-6}$ & 2,763 & 0.124 & 473,302 & 9.00 $\times 10^{-3}$& 0.96 \\
Hebrew & $1 \times 10^{-5}$ & 70 & 0.011 &  6,395 & 3.49 $\times 10^{-2}$& 1.00 \\
\hline
\end{tabular}
\label{TableSummary3}
\end{table}

\begin{table}
\caption{  Summary of {\it Google} corpus data. Annual growth rates correspond to data in the 209-year period
1800--2008.}
\begin{tabular}{@{\vrule height 10.5pt depth4pt  width0pt}lc|c|c|c|c||c|c|c|} &\multicolumn5l{Annual use $u_{i}(t)$
1-gram data}&\multicolumn3l{Annual growth $r(t)$ data}\\
\noalign{
\vskip-11pt} Corpus,\\
\cline{2-9}
\vrule depth 6pt width 0pt (1-grams)&  $N_{u} (uses)$  & $Y_{i}$ & $ Y_{f}$ & $N_{w} (words)$ & $Max[u_{i}(t)]$ & $N_{r}
(values)$ & $\langle r \rangle$ & $\sigma[r]$\\
\hline 
English  &           $3.60 \times 10^{11}$ &  1520 &  2008 &  7,380,256 &   824,591,289 &    310,987,181 &   $2.21\times
10^{-2}$ &  $0.98$\\
English fiction & $8.91 \times 10^{10}$ &  1592 &  2009 &  2,612,490 &   271,039,542 &     122,304,632 &   $2.32\times
10^{-2}$ &  $1.03$\\
Spanish &           $4.51 \times 10^{10}$ &  1532 &  2008 &  2,233,564 &   74,053,477   &     111,333,992 &  
$7.51\times 10^{-3}$ &  $0.91$\\
Hebrew &           $2.85 \times 10^{9}$ &  1539 &  2008 &  645,262 &   5,587,042   &     32,387,825 &  
$9.11\times 10^{-3}$ &  $0.90$\\
\hline\\
\end{tabular}
\label{TableSummary}
\end{table}

\end{widetext}

\end{document}